\documentclass[numberedappendix,iop]{emulateapj}
\usepackage[varg]{txfonts}
\newcommand{\se}[1]{\mbox{\S\ \ref{sec:#1}}}

\newcommand{\sesto}[2]{\S\ \ref{sec:#1}--\ref{sec:#2}}

\newcommand{\eq}[1]{Equation\ (\ref{eq:#1})}
\newcommand{\eqp}[1]{\mbox{Equation\ (\ref{eq:#1})}}
\newcommand{\eqs}[2]{Equations\ (\ref{eq:#1}) and (\ref{eq:#2})}
\newcommand{\eqsto}[2]{Equations\ (\ref{eq:#1})--(\ref{eq:#2})}

\newcommand{\Eq}[1]{Equation\ (\ref{eq:#1})}

\newcommand{\fg}[1]{\mbox{Fig.\ \ref{fig:#1}}}

\newcommand{\Fg}[1]{\mbox{Figure\ \ref{fig:#1}}}
\newcommand{\Tb}[1]{\mbox{Table\ \ref{tab:#1}}}
\newcommand{\app}[1]{\mbox{Appendix\ \ref{app:#1}}}
\newcommand{\vs}{vs.}
\newcommand{\ie}{i.e.,}
\newcommand{\eg}{e.g.,}
\newcommand{\etc}{etc}

\newcommand{\sumi}{(i)}
\newcommand{\sumii}{(ii)}
\newcommand{\sumiii}{(iii)}
\newcommand{\sumiv}{(iv)}
\newcommand{\sgn}[1]{\,\mathrm{sgn}(#1)}
\newcommand{\besselarg}{\left(\frac{2}{3}\right)}

\newcommand{\Diff}[1]{D[\Delta v_{#1}]}
\newcommand{\Diffs}[1]{D[(\Delta v_{#1})^2]}
\newcommand{\dd}{\mathrm{d}}

\shorttitle{Planetesimal-driven migration rates}
\shortauthors{Ormel, Ida, \&, Tanaka}

\begin{document}


\title{Migration rates of planets due to scattering of planetesimals}


\author{C.W. Ormel\altaffilmark{1}}
\affil{Astronomy Department, University of California,
    Berkeley, CA 94720}
\email{ormel@astro.berkeley.edu}

\author{S. Ida}
\affil{Department of Earth and Planetary Sciences, Tokyo Institute
of Technology, Meguro-ku, Tokyo, 152-8551, Japan}
\email{ida@geo.titech.ac.jp}

\author{H. Tanaka}
\affil{Institute of Low Temperature Science, Hokkaido University, Sapporo 060-0819, Japan}
\email{hide@lowtem.hokudai.ac.jp}


\altaffiltext{1}{Hubble Fellow}


\begin{abstract}
Planets migrate due to the recoil they experience from scattering solid (planetesimal) bodies.
To first order, the torques exerted by the interior and exterior disks cancel, analogous to the cancellation of the torques from the gravitational interaction with the gas (type I migration).  Assuming the dispersion-dominated regime and power-laws characterized by indices $\alpha$ and $\beta$ for the surface density and eccentricity profiles, we calculate the net torque on the planet.  We consider both distant encounters and close (orbit-crossing) encounters.  We find that the close and distant encounter torques have opposite signs with respect to their $\alpha$ and $\beta$ dependences; and that the torque is especially sensitive to the eccentricity gradient ($\beta$).  Compared to type-I migration due to excitation of density waves, the planetesimal-driven migration rate is generally lower due to the lower surface density of solids in gas-rich disk, although this may be partially or fully offset when their eccentricity and inclination are small. 
Allowing for the feedback of the planet on the planetesimal disk through viscous stirring, we find that under certain conditions a \textit{self-regulated} migration scenario emerges, in which the planet migrates at a steady pace that approaches the rate corresponding to the one-sided torque. If the local planetesimal disk mass to planet mass ratio is low, however, migration stalls. We quantify the boundaries separating the three migration regimes. 
\end{abstract}





\section{Introduction}
Bodies immersed in gaseous or particle disks migrate radially. Very small particles, strongly coupled to the gas, are carried by the gas. Thus, they follow the accretion flow or are dispersed by turbulent motions \citep{Ciesla2009,Ciesla2010}.  Larger particles tend to move on Keplerian orbits.  However, in protoplanetary disks the gas is partially pressure-supported, which causes solids to drift inwards due to the headwind they experiences \citep{AdachiEtal1976,Weidenschilling1977}.  This effect peaks for $\sim$m-size bodies (or their aerodynamic equivalents) at which they spiral in in as little as $\sim$100 orbital periods.  Larger, km-size bodies (planetesimals) are more resistant against drag-induced orbital decay due to their large inertia.  The motions of these bodies will be predominantly determined by gravitational encounters, rather than gas drag.

The gravitational interaction with the gas also causes a drag force on the planet.  The picture here is that of a massive body gravitationally perturbing the disks, which causes an excess density structure that backreacts on the planet.  One can regard the force that the planets experiences a manifestation of \textit{dynamical friction} -- a concept that is perhaps more familiar with collisionless systems, but which can also be applied to gaseous disks \citep{Ostriker1999,KimKim2007,KimKim2009,MutoEtal2011,LeeStahler2011}. When the planet is small, the resulting migration from the gravitational interaction with the gas is known as type I \citep{GoldreichTremaine1980,Ward1986}. Like dynamical friction, the type-I migration rate increases linearly with mass.  Although rather insignificant for planetesimals, it becomes very efficient for Earth-mass planets resulting in migration timescales as short as $10^5$ yr at 1 AU \citep{TanakaEtal2002}. 

For these reasons (gas-driven) migration is often invoked to explain the existence of close-in, Neptune- and Jupiter-mass planets (`hot Jupiters'), since conditions very close to the star are thought to be ill-suited to form giant planets \textit{in situ} \citep{IkomaHori2012}.  However, a clear understanding of type-I migration is somewhat complicated by the fact that it is a higher order effect; that is, the net torque on the planet results from a near cancellation of two large but opposite torques, corresponding to the respective contributions from the inner and outer disks.  In addition, the net co-orbital and Lindblad torque may have different signs. As a result the sign of type-I migration is very sensitive to the local distribution of matter, which in turn is determined by the thermodynamic properties of the disk \citep[\eg][]{PaardekooperMellema2006}.


Similar to `gas-driven' migration, scattering of solid bodies also causes a planet to migrate. This effect of planetesimal-driven migration (PDM) has been mostly explored through $N$-body studies \citep{HahnMalhotra1999,KirshEtal2009,BromleyKenyon2011,CapobiancoEtal2011}.  In some cases, these authors found an migration instability, at which the planet migrates at a rate determined by the one-sided torque \citep{IdaEtal2000}.  Under these conditions, PDM is fast.

Other studies have investigated the embryo-planetesimal interaction analytically \citep[\eg][]{Ida1990,IdaMakino1993,TanakaIda1996,Rafikov2003iii}.  Mostly, these studies consider the effect of the embryo on the planetesimal disk, \eg\ the rate at which the protoplanet excites the planetesimal's eccentricity or how it opens a gap by scattering.  

In this paper, on the other hand, we will study the \textit{recoil} of the scattering on the planet for given planetesimal properties.  These calculations provide, for the first time, an analytic expression for the two-sided torque for planetesimal scattering -- the analogue to the type-I migration torque.

We assume the following: \sumi\ a smooth disk where the spatial distribution of surface density and eccentricity are power-laws; \sumii\ the dispersion-dominated regime (relative velocities are given by the eccentricity of the planetesimals at close encounter); \sumiii\ Keplerian orbits for the planetesimals; \sumiv\ a circular orbit for the planet.  
We account for both distant and close encounters, corresponding to orbits that do or do not cross the planet (see \fg{disk}).  We then compute the recoil of the planet due to scatterings with planetesimals on orbits both interior and exterior to the planet, which results in the PDM rate.

PDM can be divided into three regimes, depending on the ratio of the planet mass compared to the mass of the solids with which it interacts: 
\begin{enumerate}
  \item Low mass planets. They do not exert a (strong) feedback on the disks. Correspondingly, the gradients in planetesimal's eccentricity and surface density are those of the background disk ($\alpha$ and $\beta$ in \fg{disk}) and can be assumed fixed during the migration; 
  \item Massive planets. They have difficulty to migrate over large distances due to their inertia.  Instead, the planet scatters away the planetesimals, leaving a gap \citep{Rafikov2003iii,Rafikov2003}. 
  \item Intermediate-mass planets. They exert some feedback on the disk but not enough to halt their migration. 
\end{enumerate}
In \sesto{model}{torques} the first regime is assumed.  In \se{model} the calculation for the migration rate due to distant and close encounters are presented.  In \se{torques} the net torque and the corresponding migration timescale are computed and compared to the type-I migration timescale. Furthermore, the approach is sketched how a distribution in eccentricity must be incorporated.  In \se{diff} the importance of diffusive motions (`noise') is investigated.  Then, in \se{sus-migr} we focus on the third regime and find that the migrating planet regulates the local eccentricity profile.  Furthermore, we will outline the boundaries dividing the regimes and find that the intermediate regime covers a large region of the parameter space.  We summarize our results in \se{diss}.


\begin{figure}[t]
  \plotone{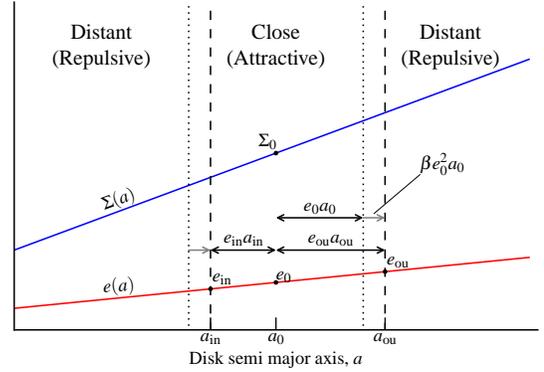}
  \caption{Sketch of the disk profile.  A planet on a circular orbit ($e=0)$ at semi-major axis $a_0$ interacts with planetesimals either through close or distant encounters.  Planetesimals that are able to cross the planet's semi-major axis interact via close encounters; otherwise the encounters are distant. We allow for a power-law profile of surface density and eccentricity with indices $\alpha$, $\beta$ (see \eq{sige0}) and compute the net migration rate $da_0/dt$ that the planet experiences due to scattering of the planetesimals in the dispersion-dominated regime. A nonzero $\beta$ causes the transition between the regimes (dotted and dashed vertical lines) to shift by an amount $\approx$$\beta e_0^2 a_0$ (see text).}
  \label{fig:disk}
\end{figure}

\begin{deluxetable}{lp{7cm}}
  \tablecaption{\label{tab:constants}List of frequently-used symbols.}
  \tablehead{
  Symbol  & Description
  }
  \startdata
  $\Delta v_i$      & Change in the $i$-th component of the relative velocity \\
  $\Gamma$          & Dimensional torque on planet\\
  $\Lambda$         & Coulomb factor \\
  $\Sigma(a),\Sigma_0$ & Surface density of planetesimals at disk radius $a$ or at reference radius $a_0$ \\
  $\Omega_0,\Omega_a$ & Orbital frequency at corresponding to the reference position or to semi-major axis $a$ \\
  $\alpha$          & Exponent in surface density power-law (\eqp{sige0})\\
  $\beta$           & Exponent in eccentricity power-law (\eqp{sige0}) \\
  $\gamma_X$        & Dimensionless torque for close or distant encounters (\eqp{gam-X}) \\
  $\gamma_\mathrm{cv}$ & Curvature component of dimensionless torque (\eqp{gam-X}) \\
  $\gamma_\nabla$   & Gradient component of dimensionless torque (\eqp{gam-X}) \\
  $\delta$          & Exponent in inclination power-law \\
  $\phi$            & Azimuthal coordinate in cylindrical coordinate system \\
  $\nu$             & Viscosity or diffusion rate in semi-major axis (\eqp{nu-scat}) \\
  $\theta$          & True anomaly ($\theta=0$ indicates periapsis; Appendix) \\
  $\Diff{i}$        & Diffusion coefficient: rate of change in relative velocity (\eqp{Diffi})\\
  $G_N$             & Newton's gravitational constant \\
  $M_\star$         & Stellar mass \\
  $M_p$             & Planet mass \\
  $Q_\mathrm{pd}$   & Combination of $q_d$ and $q_p$, \eq{Qpd} \\
  $R$               & Radial coordinate in cylindrical units \\
  $R_h$             & Hill radius \eqp{Rhill} \\
  $\tilde{P}_{Rz}$  & Probability density of finding a planetesimal near $R=a_0$ and $z=0$ (\eqp{PRz-tilde})\\
  $T_\mathrm{migr}$ & Timescale to migrate globally over distance $a_0$ \\
  $T_\mathrm{migr}^\ast$ & Timescale to migrate locally over distance $e_0 a_0$ \\
  $T_\mathrm{syn}$  & Synodical period \\
  $T_\mathrm{type-I}$ & Type-I migration timescale \\
  $T_\mathrm{vs}$   & Viscous stirring timescale  (\eqp{Tvs-0})\\
  $V_{k0}$          & Kepler (orbital) velocity corresponding to $a_0$ \\
  $a_\mathrm{[in,ou]}$  & Inner or outer-most semi-major axis from where planetesimals cross the planet's orbit \\
  $a_0$             & Semi-major axis of the planet; reference radius \\
  $b$               & Distance between semimajor axis planet and planetesimal \\
  $[e,e_0]$         & Eccentricity of planetesimals (at $a=a_0$) \\
  $e_h$             & Hill eccentricity (\eqp{eh}) \\
  $e_h^\star$       & Lower range of the Hill eccentricity for which self-regulated migration applies (\eqp{ehast}) \\
  $i$               & Inclination of planetesimals \\
  $f_\Lambda$       & Coulomb term, $f_\Lambda = \log (1+\Lambda^2)$ \\
  $g_\Lambda$       & Coulomb term, $g_\Lambda = \Lambda^2/(1+\Lambda^2)$\\
  $m$               & Mass of individual planetesimal \\
  $q_d$             & Dimensionless `disk mass' (\eqp{qdisk}) \\
  $q_p$             & Dimensionless mass of the planet (=$M_p/M_\star$) \\
  $r$               & Radial coordinate in polar coordinate system \\
  $v$               & Relative velocity between planet and planetesimal at $a=a_0$ \\
  $v_i$             & Relative velocity of $i$-th component \\
  $z$               & Vertical coordinate in cylindrical units
  \enddata
\end{deluxetable}

\section{Calculation of the migration rate}
\label{sec:model}
\subsection{Statement of the problem and methodology}
We consider the following setup.  A planet of mass $M_p$ moves on a circular, non-inclined orbit at a reference disk radius $a_0$ in the equatorial plane.  The planet interacts gravitationally with planetesimals of mass $m\ll M_p$ that are characterized by standard Keplerian orbital elements: semi-major axis $a$, inclination $i$, eccentricity $e$, and phase angles.  The surface density of the planetesimals is given by $\Sigma(a)$, and the planetesimals are assumed to be randomly distributed in their phase angles: mean anomaly $t$, and argument of periapsis, $\omega$.  The calculations allow for gradients in $\Sigma$ and $e$; specifically, they are assumed to be a power-laws with indices $\alpha$ and $\beta$:
\begin{equation}
  \Sigma(a) = \Sigma_0 \left( \frac{a}{a_0} \right)^\alpha; \quad
  e(a) = e_0 \left( \frac{a}{a_0} \right)^\beta,
  \label{eq:sige0}
\end{equation}
where $\Sigma_0$ and $e_0$ are reference values that correspond to the surface density and eccentricity at the semi-major axis of the planet ($a=a_0$).

This configuration is sketched in \fg{disk}, where the surface density and eccentricity follow a power-law profile.  There are two ways in which the planetesimals interact with the planet -- close and distant encounters -- dependent on whether or not they are able to cross the planet's orbit.  In close encounters, the planet tends to scatter planetesimals from the exterior disk to the interior disk and vice versa. This is a dispersive process: the net separation after the scattering on average increases \citep{Rafikov2003iii}. But due to the recoil from the scattering, the planet moves in the direction of the denser planetesimal belt. Thus, from \textit{its} perspective it is attracted by the belt, although it may strongly excite the belt over the course of its migration. 

Distant encounters are repulsive, in the sense that they push the planet away from a planetesimal belt.  Since the planet is assumed to move on a circular orbit, the boundary between the distant and close encounter regimes is determined by the periapsis ($a(1-e[a])$) and apoapsis ($a(1+e[a])$) of the planetesimals' elliptical orbits. Planetesimals of semi-major axes less then $a_\mathrm{in}$ or larger than $a_\mathrm{ou}$ interact through distant encounters; planetesimals of semi-axes $a_\mathrm{in}\le a \le a_\mathrm{ou}$ through close encounters (see \fg{disk}).  Here, $a_\mathrm{in},a_\mathrm{ou}$ are given by:
\begin{equation}
  \label{eq:ainou}
  a_\mathrm{in|ou} \left[ 1 \mp e(a_\mathrm{in|ou}) \right] = a_0,
\end{equation}
where the upper sign corresponds to $a_\mathrm{ou}$, the lower to $a_\mathrm{in}$.  When \eq{sige0} is linearlized by writing $e\approx e_0 +\beta e_0(a_\mathrm{in|ou} -a_0)/a_0$, we can solve for $a_\mathrm{in|ou}$ to obtain
\begin{equation}
  a_\mathrm{in|ou} \approx a_0 \pm e_0 a_0 +e_0^2 \beta a_0.
\end{equation}
A positive $\beta$ (depicted in \fg{disk}) therefore shifts the transition between the distant and the close encounter region to larger $a$ by an amount $\approx$$\beta e_0 ^2 a_0$.

For (very) low $e$ the distinction between close and distant encounters is no longer determined by the eccentricity but by the Keplerian shear of the disk.  In the limit of zero eccentricity and inclination, the border between distant and close encounters lies at $\sim$$2.5 R_h$ \citep{Nishida1983,PetitHenon1986} where $R_h$ is the Hill radius,
\begin{equation}
  R_h = a_0 \left( \frac{M_p}{3M_\star} \right)^{1/3} = a_0 \left( \frac{q_p}{3} \right)^{1/3},
  \label{eq:Rhill}
\end{equation}
with $M_p$ the planet's mass, $M_\star$ the stellar mass, and $q_p = M_p/M_\star$.  Furthermore, in the zero-eccentricity limit planetesimals of semi-major axis similar to the planet travel on horseshoe orbits and do not enter the Hill sphere.  Interactions in this shear-dominated regime are qualitatively different from the high velocity regime, where random motions (caused by the eccentricity of the planetesimal) dominate.  Random motions start to dominate over shear motions when $eV_K \gtrsim R_h \Omega_0$ or for $e \gtrsim (M_p/3M_\star)^{1/3}$, where $V_{k0} = a_0\Omega_0$ is the Keplerian orbital velocity and $a_0$ and $\Omega_0$ the local orbital frequency.  It is sometimes convenient to express eccentricities in terms of $R_h \Omega_0$ rather than $V_k$; \ie
\begin{equation}
  \label{eq:eh}
  e_h = \frac{eV_k}{R_h \Omega_0} = e \left( \frac{q_p}{3} \right)^{-1/3},
\end{equation}
so called Hill eccentricities.  In this work it is assumed that the dispersion dominated regime applies: $e_h>1$.

\subsection{The contribution from distant encounters}
\label{sec:distant}
\citet{HasegawaNakazawa1990} have calculated the change in relative semi-major axis due to a distant encounter among two bodies (see also \citealt{HenonPetit1986}).  Assuming a uniform distribution of phase angles, the average change for the encounter becomes:
\begin{equation}
  \langle \Delta b \rangle 
  = \frac{CR_h^6}{b^5}
  \label{eq:HN90}
\end{equation}
where
\begin{equation}
  C = 54 \left( \frac{8}{27}\left[ 2K_0(2/3) +K_1(2/3) \right] \right)^2 \approx 30.1
\end{equation}
is a constant, $b=a-a_0$ the separation in semi-major axis between planet and planetesimal, and $K_\nu$ is the modified Bessel function of the second kind of order $\nu$. The encounter is always repulsive; $\langle\Delta b\rangle$ has the same sign as $b$ and, after phase-averaging, is independent of eccentricity and inclination. When we consider the interacting bodies to be a planetesimal of mass $m$ and a planet of mass $M_p \gg m$, the planetesimal will experience the largest change in its semi-major axis. But due to the recoil effect, the planet experiences a change of
\begin{equation}
  \label{eq:HN90-2}
  (\Delta b)_M 
  = -\frac{m}{m+M_p} \langle \Delta b \rangle 
  \approx -\frac{m}{M_p} \langle \Delta b \rangle.
\end{equation}

The rate at which a planet migrates due to encounters with planetesimals at distance $b$ is given by \eq{HN90-2} times the encounter rate. To first order the encounter rate for an impact parameter $b$ is $\frac{3}{2}|b|\Omega_0 \Sigma_0/m$, where we took the local value of the surface density at the planet's position and linearized the (Keplerian) shear. If the planet interacts only with planetesimals on one side of its orbits, \eg\ with planetesimals exterior to it ($b>0$), the migration rate becomes:
\begin{eqnarray}
  \label{eq:dadt-dist}
  \left( \frac{da}{dt} \right)_\mathrm{1s-di} \nonumber
  &=& -\frac{1}{M_p} \int_{e_0a_0}^\infty db\, \frac{3}{2}\Sigma_0 b\Omega_0 \langle \Delta b \rangle \\
  &=& - \frac{C\Sigma M_p a_0^3 \Omega_0}{18 M_\star^2 e_0^3}
  \approx 1.67 \frac{q_dq_p}{e_0^3} (a_0 \Omega_0),
\end{eqnarray}
(when the inner disk is considered, the sign will be positive)
where the subscript `1s--di' refers to `one-sided' and `distant encounters'.
In \eq{dadt-dist} the dimensionless disk mass $q_d$ is defined as:
\begin{equation}
  \label{eq:qdisk}
  q_d(a) 
  = \frac{\Sigma(a) a^2}{M_\star} 
  \approx 10^{-6}\ \left( \frac{\Sigma_1}{10\ \mathrm{g\ cm^{-2}}} \right) \left( \frac{M_\star}{M_\sun} \right)^{-1} \left( \frac{a}{a_1} \right)^{2+\alpha},
\end{equation}
where $a_1$ is a reference radius (say 1 AU) and $\Sigma_1$ the surface density at $a=a_1$.  Note that $q_d$ is a local quantity that depends on disk radius $a_0$.  In the outer disk, $q_d$ may be significantly larger, since ices will contribute to the solid fraction and $2+\alpha$ is typically positive.

\Eq{dadt-dist} gives the migration rate due to distant encounters for one side of the disk. This zeroth order effect scales as $\propto$$e_0^{-3}$. As the contribution from the interior disk will have the opposite sign, these terms cancel to zeroth order.

Therefore, we consider the next order contributions. These arise due to gradients in eccentricity and surface density (\fg{disk}) and due to higher order approximation of the velocity field around the planet instead of the sheering sheet approximation used in the local Hill formalism which \eq{HN90} relies on. These latter effects are referred to as curvature effects. To obtain the higher order term, we have used the formalism of linear density wave theory \citep{GoldreichTremaine1980,Ward1986,Ward1997}. These provide us with a formula for the torque density -- the torque per unit disk radius -- the planet exerts on the disk. In \app{Idist} we derive the expressions for the torque that the planet experiences. The result is:
\begin{eqnarray}
  \label{eq:Gam-1s-di}
  \frac{\Gamma_\mathrm{1s-di}}{M_p (a_0 \Omega_0)^2}
  &=& q_d q_p\frac{\mp0.84}{e_0^3} \\
  \frac{\Gamma_\mathrm{2s-di}}{M_p (a_0 \Omega_0)^2}
  &=& q_d q_p\frac{-5.7 +2.5(-\alpha +2\beta)}{e_0^2}. 
  \label{eq:gam-2s-di}
\end{eqnarray}
In these expressions $\Gamma_\mathrm{1s}$ denotes the torque the planet experiences due to one side of the disk only, where the upper sign corresponds to the exterior disk (integration over the distant encounter region where $b$ is positive) and the lower sign to the interior disk. The two-sided torque $\Gamma_\mathrm{2s}$ corresponds to contributions from both sides of the disk. As remarked, this expression is an order higher in $e_0$ than the one-sided torque -- but still significant.

The torque adds or removes angular momentum to the planet at a rate $dl_z/dt$ where $l_z = a_0^2 M_p \Omega_0$. Since $\dd l/\dd t = M_p (\dd a/\dd t) \dd(a_0^2 \Omega_0)/\dd a$ we obtain the migration rate as
\begin{equation}
  \frac{da_0}{dt}
  = \frac{2\Gamma}{M_p a_0\Omega_0},
\end{equation}
and it can be verified that \eq{Gam-1s-di} is consistent with \eq{dadt-dist}. 
When accounting for both sides of the disks, the migration rate becomes
\begin{equation}
  \left( \frac{da_0}{dt} \right)_\mathrm{2s-di} 
  \approx \frac{-11.3 +5.0(-\alpha +2\beta)q_d q_p}{e_0^2} (a_0 \Omega_0).
\end{equation}
The sign of the migration thus depends on the values of $\alpha$ and $\beta$.  A large, positive value of $\alpha$ implies that interactions with the exterior disk will dominate, which pushes the planet inwards.  Positive $\beta$ implies that $a_\mathrm{in}$ lies closer (in absolute terms) to $a_0$ than $a_\mathrm{ou}$, which tilts the balance in favor of the inner disk. However, due to the large negative value of the curvature term, the direction of migration tends to be inwards in most cases.  Finally, lower $e_0$ increases the importance of distant encounters as both $a_\mathrm{in}$ and $a_\mathrm{ou}$ move closer to $a_0$.

Distant encounters represent only one side of the medal.  Close encounters reverse the sign and have the opposite dependences on $\alpha$ and $\beta$.  For the net migration rate both must be considered.   

\subsection{The contribution from close encounters}
\label{sec:close-main}
A scattering of 2 bodies rotates the relative velocity vector $\mathbf{v}$, while preserving its absolute value.  For the migration rate it is the change in the azimuthal component of $\mathbf{v}$, $\Delta v_\phi$, that matters and this component is generally not conserved. Together with the encounter rate they determine the force that the planet experiences.

\citet{BinneyTremaine2008} have calculated the \textit{diffusion coefficients} -- the rate of change in the components of $\mathbf{v}$.  After integration over impact parameters they obtain the rate of change in the parallel component \citep[][their Eqs. L.11]{BinneyTremaine2008}:
\begin{equation}
  \label{eq:Diff-vpar}
  \Diff{\parallel} = 2\pi nv \frac{G^2_N m(M_p+m)}{v^3} f_\Lambda.
\end{equation}
where $f_\Lambda=\ln (1+\Lambda^2)$ is a Coulomb term which is assumed to be constant (\ie\ independent of velocity) in the following, $G_N$ Newton's gravitational constant, $m$ the mass of the field particles (planetesimals), $n$ the \textit{local} number density (at the planet's position), and $v=|\mathbf{v}|$ the \textit{relative} velocity between the planet and the \textit{unperturbed} (Keplerian) orbit of the planetesimals at the interaction point (sometimes called collision orbits; \citealt{TanakaIda1996}).  

\Eq{Diff-vpar} does not include the effects of the solar gravity, which can be effective to change the orbital of planetesimals during the scattering. \citet{TanakaIda1996} examined the effects of the solar gravity and found that the effect of the solar gravity cancels, after averaging over the (uniformly distributed) phase angles (\ie\ the longitudes of periapsis and ascending node). This cancellation is related to the fact that the unperturbed (\ie\ Keplerian) axisymmetric particulate disk does not exert any torque on the planet. Thus the effect of the solar gravity during scattering will cancel even in our case where non-local, \ie\ curvature terms, effects are included.

For the individual components we have (\citealt{BinneyTremaine2008}, their Equation [7.89]):
\begin{equation}
  \label{eq:Diffi}
  \Diff{i} = \frac{v_i}{v} \Diff{\parallel}.
\end{equation}
Here we consider the change in the azimuthal ($\phi$) component and further assume that $M_p\gg m$:
\begin{equation}
  \label{eq:Ddelvy}
  \Diff{\phi} 
  =  2\pi G_N^2 n M_p m f_\Lambda \frac{nv_\phi}{v^3}.
\end{equation}
\Eq{Ddelvy} is nothing else than the azimuthal force exerted on the protoplanet due to dynamical friction with the planetesimals.

However, \eq{Ddelvy} is only valid when the density $n$ and velocity field $\mathbf{v}$ are uniform.  This is certainly not the case in a Keplerian disk; particles of different semi-major axis $a$ will have a different (relative) velocity at the point where they interact with the planet, that is, at $a=a_0$.  \Eq{Ddelvy} has to be convolved over the semi-major axis.  The same holds for the density, $n$.  For particles traveling on a Kepler orbit with eccentricity $e$ and semi-major axis $a$, the density is not constant as the particle's velocity depends on its position (true anomaly).

Let us introduce the projection operator $P_{Rz\phi}$, defined such that $P_{Rz\phi}\dd R (R\dd\phi) \dd z$ gives the probability of finding a particle with orbital elements $a,e,i$ and random phase angles in the interval $[R,R+\dd R; \phi+\dd \phi; z+\dd z]$, where $(R,\phi,z)$ are cylindrical coordinates.  Clearly, $P_{Rz\phi}$ is a function of the properties of the particle ($a,e,i$) as well as the position at which it is evaluated, as given by the coordinates $(R,z,\phi$).  For the latter we assume azimuthal symmetry, $R=a_0$, and $z=0$ corresponding to the position of the planet and define a new, more specific, projection operator:
\begin{equation}
  \label{eq:PRz-tilde}
  \tilde{P}_{Rz}(a,i,e) = 2\pi a_0 P_{R\phi z}(a,i,e; R=a_0, \phi, z=0),
\end{equation}
such that $\tilde{P}_{Rz}\dd R \dd z$ gives the probability that an $(a,e,i)$-particle can be found in the equatorial plane at $a_0$ at arbitrary $\phi$.  

Using \eq{PRz-tilde} we obtain the contribution to the density $n$ from particles of semi-major axis $a$.  Since the total mass of particles in $[a,a+da]$ equals $dm=2\pi a \Sigma(a) da$:
\begin{equation}
  \label{eq:dn}
  dn 
  = \frac{1}{m} 2\pi a \Sigma(a) P_{R,\phi,z}(a;a_0,z=0) da 
  = \frac{1}{m} \tilde{P}_{Rz} \frac{a\Sigma(a)}{a_0} da.
\end{equation}
With this notation the specific force on the protoplanet, \eqp{Ddelvy}, becomes:  
\begin{equation}
  \label{eq:Diff2}
  F_\phi 
  = D[\Delta v_\phi] 
  = 2\pi G^2_N M_p  f_\Lambda \int_{a_\mathrm{in}}^{a_\mathrm{ou}} \dd{a}\, \tilde{P}_{Rz} \Sigma(a) \left( \frac{a}{a_0} \right) \frac{v_\phi}{v^3},
\end{equation}
where all quantities in the integrand are functions of semi-major axis, $a$. The integration proceeds over the close encounter region.

The vertical component of the torque exerted on the protoplanet is given by $\Gamma_z = a_0 M_pF_\phi$: 
\begin{equation}
  \label{eq:Iclose-def}
  \Gamma_\mathrm{cl}
  = F_\phi a_0 M_p
  = 2\pi G^2_N M_p^2 a_0 f_\Lambda \int \dd{a}\ \tilde{P}_{Rz} \Sigma_0 \left( \frac{a}{a_0} \right)^{1+\alpha} \frac{v_\phi}{v^3},
  \label{eq:V-close-1}
\end{equation}
where we have inserted \eqp{sige0} for $\Sigma(a)$.  
%
This integral gives the migration rate due to close encounters.  To solve it, the velocity field of the planetesimals near the protoplanet (the $v$ and $v_\phi$ terms) and $\tilde{P}_{Rz}$ must be expressed as function of $a$.  These steps are outlined in \app{Iclose}. \Eq{Iclose-def} also depends on the inclination of the particles --  a thinner disk will, for example, increases the density of particles (so that $\tilde{P}$ increases) and additionally decreases the relative velocity $v$ (since $v_z$ decreases).  These effects increase the magnitude of $\Gamma_\mathrm{cl}$ and therefore the migration rate.

In \app{Iclose} we perform the calculations for the case where $i=e/2$ -- the equilibrium solution-- and a case where $i\ll e$. We obtain, for the one-sided torques:
\begin{equation}
  \label{eq:Gam-1s-cl}
  \frac{\Gamma_\mathrm{1s-cl}}{(a_0\Omega_0)^2 M_p} 
  \approx  f_\Lambda q_d q_p
    \left\{ \begin{array}{ll}
    \displaystyle
    \frac{\pm1.1}{e_0^3}  & (i=e/2 \ll 1); \\[6mm]
    \displaystyle
    \frac{\pm1.3}{i_0e_0^2}  & (i\ll e \ll 1); \\
    \end{array} \right.
\end{equation}
(the upper sign corresponds to the torque that the exterior disk exerts on the planet; the lower to the interior disk); and for the two-sided torque:
\begin{equation}
  \label{eq:Iclose-main}
  \frac{\Gamma_\mathrm{2s-cl}}{(a_0\Omega_0)^2 M_p} 
  \approx  f_\Lambda  q_d q_p
    \left\{ \begin{array}{ll}
    \displaystyle
    \frac{-0.7 +2.0(\alpha-3\beta)}{e_0^2} & (i=e/2); \\[6mm]
    \displaystyle
    \frac{1.6 + 2.3 \left( \alpha-2\beta-\delta \right)}{i_0e_0}  & (i\ll e); \\
    \end{array} \right.
\end{equation}
where $\delta$ is the exponent of the inclination dependence with disk radius, \ie\ the inclination equivalent of $\beta$. Contrary to the distant encounter torque (\eqp{gam-2s-di}), $\Gamma_\mathrm{2s-cl}$ increases with increasing $\alpha$, since for close encounter the planet tends to move in the direction of the more massive planetesimal belt.  In addition, \eq{Iclose-main} displays a negative dependence on $\beta$, which can be understood since the cross section for encounters is largest when the relative velocity (eccentricity) is lowest.  More generally, the relative importances of the gradient terms follow directly from \eq{Iclose-def} (or even \eqp{Diff-vpar}). In it $\tilde{P}_\mathrm{Rz}$ reflects the scaleheight dependence and consequently contributes a term $-\delta$; $\Sigma(a)$ a term $\alpha$; and $v_\phi/v^3$, which is proportional to $e^{-2}$, a term $-2\beta$. Note the difference in the curvature term (the first term in \eqp{Iclose-main}) between the equilibrium case ($e=i/2$) and the thin disk case ($i\ll e$): in the latter it is positive, whereas in the former it is negative.  However, the curvature term for distant encounters (\eqp{gam-2s-di}) is always negative.

Thus, the planet is attracted towards quiescent and massive planetesimal belts, in line with previous studies \citep{TanakaIda1999,PayneEtal2009,CapobiancoEtal2011}. During its migration, the planet scatters away many planetesimals, which in turn affects their spatial and dynamical distribution. The amount with which the scattering changes the (effective) values of $\alpha$ and $\beta$, depends on the relative masses of the planetesimal belt and the planet (see \se{sus-migr}).

In the remainder, we will focus on the equilibrium solution ($i/e=2$ and $\delta=\beta$) since this is the expected ratio for the dispersion-dominated regime \citep{IdaEtal1993}. We consider an extension to an eccentricity distribution in \se{Rayleigh}.


\section{Torques and migration rates}
\label{sec:torques}
\subsection{Outwards or inwards}
\begin{figure}
  \centering
  \includegraphics[width=85mm]{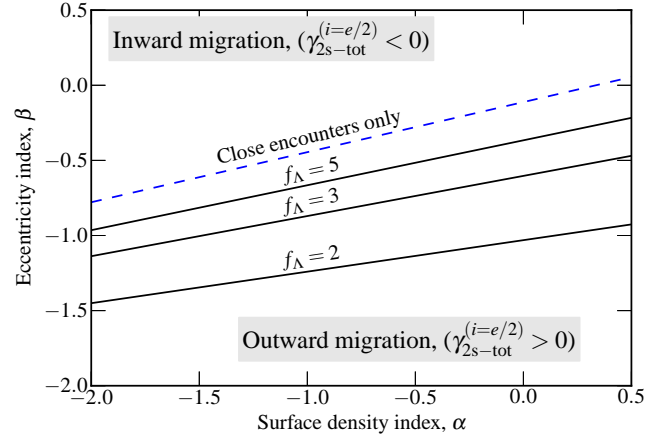}
  \caption{Sign of migration due to scattering by close and distant encounters for the two-sided torque with $i=e/2$. We plot the line in the ($\alpha,\beta$) plane (referring to the indices in surface density and eccentricity)  where the migration direction changes sign.  For close encounters migration is outwards below the blue, dashed line.  When distant encounters are included (\textit{black} lines) the dividing line shifts down, dependent on the value of the Coulomb factor $f_\Lambda$.}
  \label{fig:Itot}
\end{figure}

\begin{deluxetable*}{llrrrrl}
  \tabletypesize{\small}
  \tablewidth{\textwidth}
  \tablecaption{\label{tab:torques}Dimensionless torques for planetesimal-driven migration $\gamma$.}
  \tablehead{
  Torque type                           & Symbol        & Leading term                        & Zeroth-order term           &  Curvature term                       & \multicolumn{2}{c}{Gradient terms} \\
                                        & $\gamma_X$              & $F(e_0,i_0)$                         &                             &  $\tilde\gamma_\mathrm{cv}$                 & $\tilde\gamma_\nabla$    & $\alpha+g(\beta,\delta)$ \\
  (1)                                   & (2)             & (3)                                  &  (4)                        & (5)                                   & (6)    & (7) }
  \startdata
  Distant, 1-sided                      & $\gamma_\mathrm{1s-di}$     & $e_0^{-3}$                    & $\mp0.836$ \\
  Distant, 2-sided                      & $\gamma_\mathrm{2s-di}$     & $e_0^{-2}$                    &                        & $-5.66$                                                          & $-2.51$ & $(\alpha-2\beta)$ \\
  Close, 1-sided, $i=e/2$               & $\gamma_\mathrm{1s-cl}^{(i=e/2)}$ & $e_0^{-3}$              & $\pm1.14f_\Lambda$ \\
  Close, 1-sided, $i\ll e$              & $\gamma_\mathrm{1s-cl}^{(i\ll e)}$ & $i_0^{-1}e_0^{-2}$     & $\pm1.27f_\Lambda$ \\
  Close, 2-sided, $i=e/2$               & $\gamma_\mathrm{2s-cl}^{(i=e/2)}$     & $e_0^{-2}$          &                        & $-0.66f_\Lambda$                                                 & $1.97f_\Lambda$ & $(\alpha -3\beta)$ \\
  Close, 2-sided, $i\ll e$              & $\gamma_\mathrm{2s-cl}^{(i\ll e)}$ & $i_0^{-1}e_0^{-2}$     &                        & $1.63f_\Lambda$                                                  & $2.28f_\Lambda$ & $(\alpha -2\beta-\delta)$

  \enddata
  \tablecomments{Summary of dimensionless torques expression for planetesimal scattering in the dispersion-dominated regime (see \eqp{gam-X}).  Column (1): torque: distant or close; one or two sided; and the inclination model. Column (2): corresponding symbol. Column (3): the order of the contribution to the torque in terms of the eccentricity and inclination at the reference position (see \eqp{gam-X}). Column (4): the zeroth order contribution from one side of the disk (the upper sign corresponds to the outer disk; the lower to the inner); Column (5): the contribution to $\gamma$ that arises due to curvature, \ie\ the deviation from the linear approximation; Column (6) and (7): the contribution to $\gamma$ due to gradients in surface density and eccentricity (see \fg{disk}).}
\end{deluxetable*}

Let us decompose the dimensional torque $\Gamma_X$ for an interaction $X$ as follows:
\begin{eqnarray}
  \label{eq:gam-X}
  \frac{\Gamma_X}{M_p (a_0\Omega_0)^2}
  &\equiv&  q_d q_p F_X(e_0,i_0) \times \gamma_X(\alpha,\beta,\delta) \\
  &\equiv&  q_d q_p F_X(e_0,i_0) \times \left[ \tilde\gamma_\mathrm{cv} +\tilde\gamma_\nabla (\alpha +g(\beta,\delta)) \right].
\end{eqnarray}
Here, $F_X(e_0,i_0)$ is a function of $e_0$ and $i_0$ only (without a numerical prefactor) and $\gamma_X$ is the dimensionless torque for interaction $X$. For a two sided toques, $\gamma$ is further decomposed into a curvature term $\tilde\gamma_\mathrm{cv}$ and a gradient term $\tilde\gamma_\nabla$, defined such that it is proportional to $\alpha$ in $\gamma_X$. 
For example, for the two-sided, close encounter torque of \eq{Iclose-main}, $\gamma_\mathrm{2s-cl}^{(i=e/2)} = (-0.7 +2.0[\alpha-3\beta])f_\Lambda$, $F= e_0^{-2}$, $\tilde\gamma_\mathrm{cv}=-0.7f_\lambda$, $\tilde\gamma_\nabla=2.0f_\lambda$ and $\alpha=-3\beta$. We have compiled a list of dimensionless torques in \Tb{torques}.


The direction of the migration is determined by the sign of $\gamma_\mathrm{tot} = \gamma_\mathrm{cl} +\gamma_\mathrm{di}$ and is positive for outwards migration, negative for inwards migration. It depends on $\alpha$, $\beta$ and on the value of the Coulomb term $f_\Lambda = \log (1+\Lambda^2)$.  For example:
\begin{equation}
  \gamma^{(i=e/2)}_\mathrm{2s-tot}(\alpha,\beta,f_\Lambda) \approx \left\{ \begin{array}{ll}
   -6.3 -0.5\alpha  -0.9\beta & (f_\Lambda=1) \\
   -7.6 +3.4\alpha -12.7\beta & (f_\Lambda=3)\\
   -8.9 +7.3\alpha   -24\beta & (f_\Lambda=5) \\
  \end{array} \right.
\end{equation}
Thus, for increasing $f_\Lambda$, $|\gamma_\mathrm{tot}|$ generally becomes larger with the sign of the migration more likely to be determined by that of the close encounter contribution.  This is illustrated in \fg{Itot}.  For the range in $\alpha$ and $\beta$ displayed in \fg{Itot}, $\gamma_\mathrm{di}$ is always negative (inward migration), mainly due to the large (negative) value of the curvature term. Close encounters more readily give rise to outward migration, but require a more massive outer disk (large $\alpha$) and/or a sufficiently low eccentricity gradient (reflecting a dynamically colder state with which the planet interacts more strongly).  Accounting for both distant and close interactions shifts the boundary line dividing the inward and outward migration regime down. The amount of the shift depends on the Coulomb parameter $f_\Lambda$, which therefore translates in a relative measure of the importance of close \vs\ distant interactions.


What is the expected value of $f_\Lambda = \log(1+\Lambda^2)$?  Here, $\Lambda \simeq b_\mathrm{max}/b_\mathrm{90}$ is the ratio for the largest and typical impact radii \citep{BinneyTremaine2008}.  For the former we may substitute the disk scaleheight, $a_0 i_0$ while the latter -- the impact radius that causes a $\pi/2$ change in the relative velocity after the scattering -- is, in the high velocity regime, $b_{90}=G_N M_p/(eV_{k0})^2 = 3R_h/e_h^2$ where $e_h$ is the Hill eccentricity (\eqp{eh}).  Assuming the dispersion-dominated regime, $i=e/2$, $f_\Lambda \simeq \log (1+e_h^3/6)$.  For $e_h \gtrsim 3$, $f_\Lambda \gtrsim 3$ and the close encounter contribution typically determines the sign.  But at small Hill eccentricity $f_\Lambda \simeq 1$ and distant encounters become more important (see \fg{Itot}).

Assuming that the random motion of planetesimals is balanced by gas drag, it follows that the Hill eccentricity $e_h$ is independent of the planet mass, and has a weak ($\propto$$X^{1/5}$) dependence on the planetesimal radius and gas density (smaller planetesimals or denser gas results in lower $e_h$), disk radius (larger $a_0$ have lower $e_h$).  A typical range may be $e_h\simeq3$--8 \citep{KokuboIda2002}.  When the gas is absent, $e_h$ will increase with time until it is equilibrated by collisional damping.

\subsection{The migration timescale}
The migration timescale is defined:
\begin{equation}
  \label{eq:T-migr-X}
  T_\mathrm{migr}
  = \left( \frac{1}{a_0}\frac{da_0}{dt} \right)^{-1}
  = \left( \frac{2\Gamma_X}{M_p a_0^2 \Omega_0} \right)^{-1}.
\end{equation}
In terms of $\gamma_\mathrm{tot}$ (\eq{gam-X}) the timescale corresponding to the two-sided, $i=e/2$ torque becomes:
\begin{eqnarray}
  \label{eq:Tmigr}
  T_\mathrm{migr}
  &=& \frac{e_0^2}{2\gamma_\mathrm{tot} q_d q_p} \Omega_0^{-1} \\ \nonumber
  &\approx& 2\times 10^5 \left(\frac{e_0}{0.02}\right)^2 \left|\frac{\gamma_\mathrm{tot}}{10}\right|^{-1} \left( \frac{q_d}{10^{-4}} \right)^{-1} \left( \frac{q_p}{10^{-6}} \right)^{-1} \Omega_0^{-1}.
\end{eqnarray}
Alternatively, the migration timescale can be expressed in terms of Hill eccentricity $e_h$ using \eq{eh}:
\begin{eqnarray}
  \label{eq:Tmigr-hill}
  T_\mathrm{migr} 
  &=& \frac{e_h^2}{3^{2/3}\gamma_\mathrm{tot}q_d q_p^{1/3}} \Omega_0^{-1} \\ 
  &\approx& \nonumber
  2.2\times10^5 \left(\frac{e_h}{3}\right)^2 \left|\frac{\gamma_\mathrm{tot}}{10}\right|^{-1} \left( \frac{q_d}{10^{-4}} \right)^{-1} \left( \frac{q_p}{10^{-6}} \right)^{-\frac{1}{3}} \Omega_0^{-1},
\end{eqnarray}
which is useful since the expressions are valid only for $e_h>1$.  For reference, we also give the one-sided migration timescale due to close encounters corresponding to $\gamma_\mathrm{1s-cl}^{(i=e/2)}$:
\footnote{\Eq{T1s} is consistent with Equations (19) and (20) of \citet{IdaEtal2000}. On the other hand, Equation (23) of \citet{IdaEtal2000} is inconsistent with \eq{T1s} by a large factor ($\sim$10--100), since several numerical constants were omitted.}
\begin{eqnarray}
  \label{eq:T1s}
  T_\mathrm{1s} 
  &\approx& \frac{1}{2.2f_\Lambda} \frac{e_0^3}{q_d q_p} \Omega_0^{-1} \\ \nonumber
  &\approx&  1.1\times10^4 \left( \frac{f_\Lambda}{3} \right)^{-1} \left( \frac{e_0}{0.02} \right)^3 \left( \frac{q_d}{10^{-4}} \right)^{-1} \left( \frac{q_p}{10^{-6}} \right)^{-1} \Omega_0^{-1},
\end{eqnarray}

\subsection{Comparison with type I migration}
\label{sec:type-I}
The migration timescale for type I migration is given by \citet{TanakaEtal2002}: 
\begin{eqnarray}
  \label{eq:Ttype-I}
  T_\mathrm{type-I} 
  &=& \frac{1}{2\gamma_I(p)} \left( \frac{c_s}{V_{k0}} \right)^2 \left( \frac{\Sigma_\mathrm{g,0} a_0^2}{M_\star} \right)^{-1} \left( \frac{M_p}{M_\star} \right)^{-1} \Omega_0^{-1} \\ \nonumber
  &=& 2.5\times10^4\ \left( \frac{\gamma_I(\alpha)}{5} \right)^{-1} \left( \frac{c_s/V_k}{0.05} \right)^2 \left( \frac{q_g}{10^{-2}} \right)^{-1} \left( \frac{q_p}{10^{-6}} \right)^{-1} \Omega_0^{-1},
\end{eqnarray}
where $\gamma_I$ denotes the dimensionless torque for type-I migration, and $q_g=\Sigma_g a_0^2/M_\odot$ the dimensionless disk mass in gas.  The expression of $\gamma_I$ depends on several factors, including the type of torque considered (co-orbital, Lindblad), the surface density exponent, the pressure exponent, and the exponent for the scaleheight (see \citet{TanakaEtal2002} and successor works, \eg\ \citealt{D'angeloLubow2010,PaardekooperEtal2010,PaardekooperEtal2011,Masset2011}). Generally, the situation for type-I is more complex, because of the pressure effects and the energy transfer in gaseous disks.
Nevertheless, the similarity between \eqs{Tmigr}{Ttype-I} is striking.  In fact \eqs{Tmigr}{Ttype-I} \textit{are} the same, except that the eccentricity in \eq{Ttype-I} has been substituted by $c_s/V_\mathrm{k0}$ and the surface density in solids ($\Sigma_0$) by that of the gas ($\Sigma_{g,0}$).

\subsection{The effect of an eccentricity distribution}
\label{sec:Rayleigh}
Up till now we have neglected an intrinsic probability distribution in eccentricity and inclination, which may be expected from gravitationally-interacting bodies \citep{IdaMakino1992,OhtsukiEmori2000}. Specifically, the Rayleigh distribution
\begin{equation}
  P_R(e|\sigma_e) 
  = \frac{2e}{\sigma_e^2} \exp\left[ -\left( \frac{e}{\sigma_e} \right)^2 \right],
  \label{eq:PRayl-e}
\end{equation}
is often considered, where $\sigma_e$ is the rms-value of the eccentricity. Naively, one might expect that the distribution-averaged torque is just the torque expressions calculated listed in \Tb{torques} averaged by $P_R(e_0|\sigma_{e0})$. However, this implies that the eccentricity distribution at a distance $b$ is merely shifted in $e$, which is \textit{not} the same as a gradient in $\sigma_{e}$ -- the situation we consider here.

Thus, we write $\sigma_e \approx \sigma_{e0} +\beta b \sigma_{e0}$, where $\beta$ is the gradient with respect to $\sigma_e$ and $b$ the dimensionless separation $b=(a-a_0)/a_0$. For $b\ll1$ we can expand \eq{PRayl-e} with respect to $b$ and write $P_R(e|\sigma_e) \approx P_R(e|\sigma_{e0}) +\beta b P_C(e|\sigma_{e0})$, where $P_C$ is a correction term:
\begin{equation}
  P_C (e|\sigma_{e0}) 
  = 4e \frac{e^2 -\sigma_{e0}^2}{\sigma_{e0}^4} \exp\left[ -\frac{e^2}{\sigma_{e0}^2} \right].
\end{equation}
\begin{figure}
  \plotone{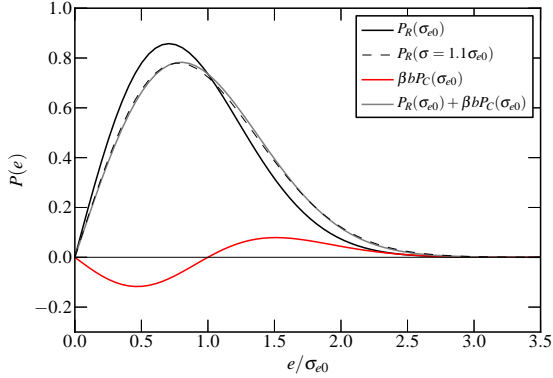}
  \caption{Eccentricity distributions: (black curve) Rayleigh distribution for $\sigma=\sigma_{e0}$; (dashed curve) Rayleigh distribution for $\sigma=1.1\sigma_{e0}$, which corresponds to the distribution at separation $b=0.1/\beta$ in the case where $\sigma_e$ is a power-law with exponent $\beta$; (red curve) correction distribution $\beta bP_C$; (gray curve) approximation to $P_R(\sigma[b])$ in terms of $P_R(\sigma_{e0})$ and $P_C(\sigma_{e0})$.}
  \label{fig:Rayleigh}
\end{figure}
The situation is illustrated in \fg{Rayleigh}. At the reference point $a_0$ the velocity distribution is characterized by an rms-value $\sigma_{e0}$. For positive $\beta$ the rms-value increases. The thin dashed line in \fg{Rayleigh} illustrates the case where $\sigma_e$ has increased by 10\% to $1.1\sigma_{e0}$, which corresponds to a distance $b=0.1/\beta$. Compared to the eccentricity distribution at $b=0$, the distribution at $b=0.1/\beta$ has an excess of high eccentricity planetesimals and a deficit of low eccentricity planetesimals. To first order in $b$ this change is represented by $\beta b P_C(\sigma_{e0})$. As a first order approximation, therefore, the distribution at $b$ is well approximated by the superposition of $P_R(\sigma_{e0})$ and $\beta b P_C(\sigma_{e0})$.

The $\beta b P_C$ component is linear in $b$.  The integration of this term over $b$ thus behaves the same as the integration of the gradient in the surface density. To see this, recall that we linearized $\Sigma=\Sigma_0(a/a_0)^\alpha$ as $\Sigma \approx \Sigma_0 (1 +\alpha b)$ and found that the $\alpha b$ term gave rise to a torque $\alpha\tilde\gamma_\nabla$ (see \se{torques}). Analogously, integration of $\beta b P_C$ will yield a term $\beta \tilde\gamma_\nabla P_C$ in the expression for the dimensionless torque. The integration of the $P_R$ component proceeds as before, except that $\beta=0$ since the eccentricity gradient is already included via $P_C$. Having accounted for the spatial distribution in this way, the resulting expression must yet be averaged over the Rayleigh velocity distribution at $a_0$. Without loss of generality, we will consider a two-sided torque for which $F(e_0)=e_0^{-2}$ (see \Tb{torques}); the distribution-averaged torque then reads:
\footnote{Furthermore, when the close encounter torque is considered, a term $-\delta \gamma_\nabla$ must be added to the terms within the square brackets \eq{Gam-torque-new}. This, to account for the scaleheight dependence. The $\delta$-term will likewise propagate into \eq{Gam-new-fin}. }
\begin{eqnarray}
  \label{eq:Gam-torque-new}
  \frac{\langle\Gamma_{2s}\rangle}{(a_0\Omega_0)^2 M_p}
  &=& q_p q_d \times \\
  &&\int de_0 \frac{ \left[ \tilde\gamma_\mathrm{cv} +\alpha\tilde\gamma_\nabla \right] P_R(e_0|\sigma_{e0}) +\beta\tilde\gamma_\nabla P_C(e_0|\sigma_{e0}) }{e_0^2}.
  \nonumber
\end{eqnarray}
The integration diverges for $e_0\rightarrow0$, which is because the shear-dominated regime is not covered in this work. Therefore the integration is cut off at the point where the Hill velocity $e_h$ becomes 1 (see \eq{eh}). We find:
\footnote{The integrals in \eq{Gam-torque-new} evaluate to
\begin{eqnarray}
  \int_{e_1}^\infty \frac{P_R(e_0|\sigma_{e0})}{e_0^2} de_0
  &=& \frac{\Gamma\left(0,\frac{e_1^2}{\sigma_{e0}^2}\right)}{\sigma_{e0}^2} 
  \approx -\frac{\gamma +2\log\frac{e_1}{\sigma_{e0}}}{\sigma_{e0}^2} \\
  \nonumber
  \int_{e_1}^\infty \frac{P_C(e_0|\sigma_{e0})}{e_0^2} de_0
  &=& \frac{2\exp\left[-\frac{e_1^2}{\sigma_{e0}^2}\right] -\Gamma\left(0,\frac{e_1^2}{\sigma_{e0}^2}\right)}{\sigma_{e0}^2}
  \approx \frac{2\left(1+\gamma +2\log\frac{e_1}{\sigma_{e0}}\right)}{\sigma_{e0}^2}
\end{eqnarray}
where $\Gamma(0,x)$ is here the incomplete gamma function, $\gamma\approx0.577$ the Euler-Mascheroni constant, and where we have expanded with respect to $e_1$, the lower cut-off of the integrations. In \eq{Gam-new-fin} we have only kept the logartihmic terms in the denominator and inserted $e_1=\sigma_{e0}/\sigma_{e0,h}$.
}
\begin{equation}
  \label{eq:Gam-new-fin}
  \frac{\langle \Gamma_\mathrm{2s}\rangle}{(a_0\Omega_0)^2 M_p}
  \sim (2\log \sigma_{e0,h}) q_p q_d \frac{\tilde\gamma_\mathrm{cv} +\tilde\gamma_\nabla(\alpha-2\beta)}{\sigma_{e0}^2},
\end{equation}
where $\sigma_{e0,h}$ is the rms-eccentricity expressed in Hill units. \Eq{Gam-new-fin} is very similar to the torque expressions obtained for the single-value power-laws.

The above is merely a sketch for the inclusion of a velocity distribution. It is not complete, since a restriction on the inclination ($i=e/2$ or $i\ll e$) is enforced. More realistically, the velocity distribution will be two dimensional and read, instead of \eq{PRayl-e}:
\begin{equation}
  P_R(i,e|\sigma_e,\sigma_i) 
  = \frac{4ie}{\sigma_e^2\sigma_i^2} \exp\left[ -\left( \frac{e}{\sigma_e}\right)^2 -\left( \frac{i}{\sigma_i} \right)^2 \right].
\end{equation}
In addition, we have in \eq{Gam-new-fin} neglected variations of the Coulomb factor, $f_\Lambda$, with $i$ and $e$. These may become important when accounting for a distribution average \citep{IdaEtal1993}.  Finally, for a general treatment, an expression for the torque at arbitrary $i_0$ and $e_0$ is needed. A general expression is derived in \app{general}. Therefore, the framework to (numerically) compute a truly Rayleigh-distributed average torque is in place.

\section{The role of diffusion in close encounters}
\label{sec:diff}
Encounters with single planetesimals result in either inward or outward kicks dependent on the direction of the scattering. Migration therefore always has a stochastic (random) component. However, the inward and outward contributions are not equal.  What we have calculated by the two-sided torques expressions of \se{torques} -- the residual -- is the systematic component. Which component will dominate depends on the ratio of the planetesimal mass over the planet mass, $m/M_p$, and the lengthscale of interest. 

To quantify the migration rate due to stochastic motions, we calculate the diffusion coefficient, $\Diffs{\phi}$, which follows from the change in the diffusion rates of the parallel and perpendicular components \citep{BinneyTremaine2008}:
\begin{equation}
  D[\Delta v_i \Delta v_j] = \frac{v_iv_j}{v^2} \left\{ \Diffs{\parallel} -\frac{1}{2}\Diffs{\perp} \right\} +\frac{1}{2}\delta_{ij}\Diffs{\perp}
\end{equation}
with
\begin{eqnarray}
  \Diffs{\parallel} &=& 4\pi nv \frac{G^2_N m^2}{v^2} g_\Lambda \\
  \Diffs{\perp}     &=& 4\pi nv \frac{G^2_N m^2}{v^2} (f_\Lambda -g_\Lambda)
\end{eqnarray}
where $g_\Lambda=\Lambda^2/(1+\Lambda^2)$.  Thus, in our case we must calculate
\begin{equation}
  \label{eq:Ddelvy2}
  \Diffs{\phi} = 2\pi G^2_N m^2 \left\{ \frac{nv_\phi^2}{v^3} [3g_\Lambda-f_\Lambda] +\frac{n}{v}[f_\Lambda-g_\Lambda] \right\}.
\end{equation}
We next perform similar operations as outlined in \se{close-main}.  That is, we express $n$ by the density function $\tilde{P}_{Rz}$ (\eqp{dn}) and integrate over the semi-major axis.  The steps are outlined in \app{diff}.  We obtain
\begin{equation}
  \Diffs{\phi} \approx \frac{3.5 f_\Lambda -1.5 g_\Lambda}{e_0^2} \frac{q_d q_p m}{M_p} a_0^2 \Omega_0^3 .
  \label{eq:diff-v2}
\end{equation}
Rather than the diffusion of $(\Delta v_\phi)^2$ we seek the diffusion in $a_0^2$, $D[(\Delta a_0)^2]$, which we refer to as the viscosity $\nu$.  Since $\Delta a = 2\Delta v_\phi/\Omega_0$ we have
\begin{equation}
  \nu 
  = \frac{4\Diffs{\phi}}{\Omega_0^2} 
  \approx \frac{14f_\Lambda -6g_\Lambda}{e_0^2} q_d q_p \left( \frac{m}{M_p} \right) a_0 V_{k0}.
  \label{eq:nu-scat}
\end{equation}
This is the diffusion coefficient for planets that results from the backreaction to the scattering of planetesimals.  Contrary to the migration rate it does not depend on the exponents of $\alpha$ and $\beta$, but it does involve the mass of the planetesimal ($m$). When $m\ll M_p$ there are many encounters, whose individual kicks are small, resulting in a smooth migration rate.  When $m$ starts to approach $M_p$, on the other hand, the importance of diffusive (random) motion increases.  As a result, the migration becomes increasingly `noisy'. The critical lengthscale is given by $L^\ast \sim \sqrt{\nu T_\mathrm{migr}} \sim \sqrt{m/M_p\gamma} a_0$. For $L\ll L^\ast$ diffusive behavior will dominate. On scales $L \gg L^\ast$ the migration occurs smoothly. 

\subsection{Comparison to \citet{OhtsukiTanaka2003}}
The diffusion coefficient for planets (\eqp{nu-scat}) may be compared to the coefficient applicable for an \textit{equal-mass} planetesimal swarm as calculated by \citet{OhtsukiTanaka2003} (their Equation [18]):
\begin{equation}
  \nu_\mathrm{OT03} = \frac{24 f_\Lambda I_\mathrm{RVS}(\beta)}{\pi e_h i_h} (N_s a_0^2) (2^{1/3}h_m)^4 a_0^2 \Omega,
  \label{eq:OT03}
\end{equation}
where $N_s$ is the column density, $h_M = (q_p/3)^{1/3}$ and $I_\mathrm{RVS}(\beta)\approx0.3$ for $i=e/2$.  Expressing \eq{OT03} in our notation, we find 
\begin{equation}
  \nu_\mathrm{OT03} \approx \frac{4f_\Lambda}{e_0^2} q_d q_p \frac{m}{M_p} a_0 V_\mathrm{k0}
\end{equation}
which is of the same magnitude as \eq{nu-scat}.  Thus, a planet of mass $M_p$ interacting with a swarm of planetesimals of mass $m$ diffuses at the same rate as the planetesimals do by interacting among themselves! Physically, the increase in cross section for encounters between planet and planetesimal due to its larger mass ($M_p$) is balanced by the decreasing kick ($\Delta a$) the planet receives.  The latter scales as $m/M_p$, while the cross section for encounters in the dispersion-dominated regime scales as $\sigma \propto b_{90}^2 \propto (GM_p)^2$ \citep[\eg\ ][]{BinneyTremaine2008}.  Thus, $\nu \propto (\Delta a)^2 \sigma$ stays constant.

\section{Self-regulated planet migration}
\label{sec:sus-migr}
\begin{figure}
  \includegraphics[width=85mm]{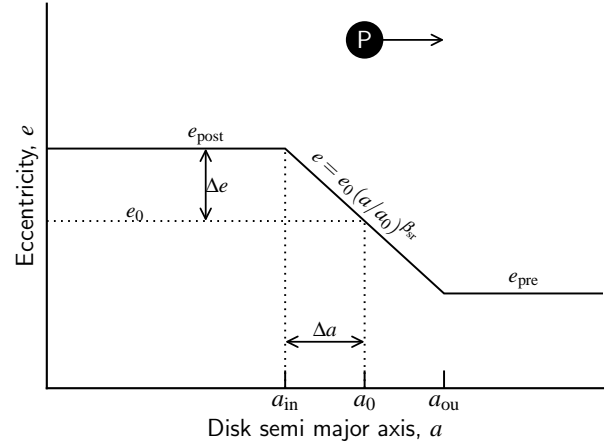}
  \caption{Sketch of eccentricity profile in the self-regulated regime. A planet at $a_0$ migrates in the direction of a less eccentric planetesimal disk (outwards in this sketch), interacting with planetesimal through close encounters over a width $\Delta a$.  During its passage, the planet slightly excites the planetesimal disk from pre-encounter eccentricities $e_\mathrm{pre}$ to post-encounter eccentricities $e_\mathrm{post}$.  We solve for the jump in eccentricity $\Delta e_0 = (e_\mathrm{post}-e_\mathrm{pre})/2$ and the corresponding eccentricity index $\beta_\mathrm{sr}$ assuming that the relative increase in eccentricity is small, $\Delta e_0/e_0 \ll1$.  However, even though $\Delta e_0/e_0 \ll1$ the eccentricity exponent $\beta_\mathrm{sr} \approx \Delta e_0/e_0^2$ can become large ($\gg$1), affecting the migration rate.}
  \label{fig:qj-sketch}
\end{figure}
The migration timescale (\eq{Tmigr}) that we have derived assumes that the local distribution of eccentricity and surface density of the planetesimals is given by the power-law indices $\alpha$ and $\beta$ that characterize the protoplanetary disk on global scales.  That is, any feedback of the planet on the disk that can change $\alpha$ and $\beta$ locally is ignored.  We will now relax this assumptions and consider the case where the protoplanet slightly excites the motions of the planetesimals, increasing their eccentricity, and altering the local eccentricity profile.

Specifically, we consider the configuration sketched in \fg{qj-sketch}, where the eccentricity profile is steady in the frame of the migrating planet.  The planet can migrate outwards or inwards (\fg{qj-sketch} depicts outwards migration).  During its passage, which is defined as the time during which planetesimals interact through close encounters, planetesimals at a pre-stirring eccentricity $e_\mathrm{pre}$ are excited to an eccentricity $e_\mathrm{post}$.  We assume that the eccentricity jump, $\Delta e \ll e_0$, where $e_0$ is the eccentricity at the reference radius.  Moreover, we assume that the eccentricity changes gradually, which implies that a planetesimal experiences many (small) encounters, before the planet has moved away from the interaction zone.  Let the half-width of the interaction zone be denoted $\Delta a \approx e_0 a_0 \ll a_0$.  Then, we can expand \eq{sige0} to obtain $e\approx e_0 + \beta e_0 \Delta a/a_0$; therefore, $\Delta e \approx \beta e_0^2$ or
\begin{equation}
  \beta \approx \frac{\Delta e}{e_0^2}.
  \label{eq:beta}
\end{equation}
This equation suggests that $\beta$ can become large, even though $\Delta e_0/e_0 \ll1$.

We seek to obtain an expression for the eccentricity jump, $\Delta e$.  We may write
\begin{equation}
  \Delta e \approx \frac{T_\mathrm{migr}^\ast}{T_\mathrm{vs}(e_0)} e_0,\quad (T_\mathrm{migr}^\ast \ll T_\mathrm{vs})
  \label{eq:Delta-e}
\end{equation}
where $T_\mathrm{migr}^\ast$ is the timescale to migrate locally over a distance $\Delta a = e_0 a_0$ and $T_\mathrm{vs}$ is the stirring timescale.  Indeed, we should have that $T_\mathrm{migr}^\ast \ll T_\mathrm{vs}$ since we assume that $\Delta e \ll e_0$.  Physically, this means that the planet migrates faster than it can excite the planetesimal disk.  The local migration timescale $T_\mathrm{migr}^\ast$ is therefore just \eq{Tmigr} multiplied by $e_0$:
\begin{equation}
  T_\mathrm{migr}^\ast = \frac{e_0^3}{12f_\Lambda \beta} q_d^{-1} q_p^{-1} \Omega_0^{-1},
  \label{eq:Tmigr-ast}
\end{equation}
where we assumed that $\gamma_\mathrm{tot} \approx \gamma_\mathrm{cl} \approx -6\beta f_\Lambda$ is entirely due to the $\beta$-dependence in $\gamma_\mathrm{cl}$.  The viscous stirring timescale in the dispersion-dominated regime is
\footnote{The viscous stirring timescale is obtained from the relaxation timescale (\citealt{Chandrasekhar1942}; see also \citealt{IdaMakino1993}):
\begin{equation}
  \label{eq:Tvs-0}
  T_\mathrm{vs} \simeq T_\mathrm{ch} = \frac{v^3}{4\pi n_p (G_N M_p)^2 \ln \Lambda},
\end{equation}
where $n_p$ is the density of perturbers.  In our case $n_p$ is the single protoplanet divided by its `stirring volume' $(2\pi a_0) \times (2e_0 a_0) \times (2 i_0 a_0)$.  For $v=e_0 a_0 \Omega_0$, $i_0=e_0/2$, $G_N M_\star = a_0^3 \Omega_0^2$ and $\ln \Lambda \approx f_\Lambda/2$, one obtains \eq{Tvs}.
}
\begin{equation}
  T_\mathrm{vs} \approx \frac{2e_0^5}{q_p^2 f_\Lambda} \Omega_0^{-1}.
  \label{eq:Tvs}
\end{equation}
With these expressions \eq{Delta-e} becomes
\begin{equation}
  \label{eq:Delta-e-2}
  \Delta e \approx \frac{q_p}{24 \beta e_0 q_d}.
\end{equation}
Equating \eq{Delta-e-2} with $\Delta e$ in \eq{beta} we obtain the eccentricity power law index for self-regulated migration:
\begin{equation}
  \label{eq:beta-2}
  \beta_\mathrm{sr}
  = \sqrt{ \frac{q_p}{24 q_d e_0^3} }
  = \sqrt{ \frac{1}{8 q_d e_h^3} }
  \approx 6.8 \left( \frac{q_d}{10^{-4}} \right)^{-1/2} \left( \frac{e_h}{3} \right)^{-3/2},
\end{equation}
where we have switched to Hill eccentricities (see \eqp{eh}).  With \eq{beta-2} we can solve for $T_\mathrm{migr}^\ast$ and the (global) migration timescale in the self-regulated regime, $T_\mathrm{sr} = T_\mathrm{migr}^\ast/e_0$:
\begin{eqnarray}
  \label{eq:Tsr}
  T_\mathrm{sr} 
  &=& \frac{e_0^{7/2}}{\sqrt{6}f_\Lambda q_d^{1/2} q_p^{3/2}} \Omega_0^{-1} \\ \nonumber
  &\approx& 1.5\times10^4\ \frac{f_\Lambda}{3} \left( \frac{e_0}{0.02} \right)^{7/2} \left( \frac{q_d}{10^{-4}} \right)^{-1/2} \left( \frac{q_p}{10^{-6}} \right)^{-3/2} \Omega_0^{-1}.
\end{eqnarray}
This much shorter timescale than \eq{Tmigr} can be attributed to the large $\beta_\mathrm{sr}$ value.  For our fiducial parameters $\beta_\mathrm{sr} \approx 7$.  Nevertheless, the eccentricity jump $\Delta e$ (\eqp{Delta-e-2}) is only $3\times10^{-3}$, much less than $e_0$.

\begin{figure}
  \plotone{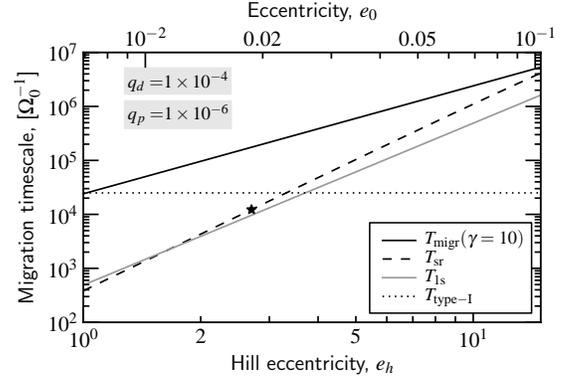}
  \caption{Migration timescales as function of eccentricity for a disk mass $q_d=10^{-4}$ and a planet mass $q_p=10^{-6}$. Shown are the migration timescale in terms of the inverse orbital frequency corresponding to: \sumi\ a fixed dimensionless torque value $|\gamma_\mathrm{tot}|=10$ (\textit{solid, black} line; \eqp{Tmigr}); \sumii\ the self-regulated migration scenario of \se{sus-migr}, where we solve for the local eccentricity gradient $\beta$ (\textit{dashed, black} line; \eqp{Tsr}); \sumiii\ the one-sided torque only (\textit{gray} line; \eqp{T1s}); \sumiv\ the type-I migration rate (\textit{dotted, horizontal} line; \eqp{Ttype-I}) using $q_g=10^{-2}$.  Our expression for $T_\mathrm{sr}$ will break down below an eccentricity $e_h^\ast$ (\eqp{ehast}), indicated by a star.}
  \label{fig:Tmigr}
\end{figure}
\Fg{Tmigr} plots the migration timescale in the self-regulated limit, \eq{Tsr}, as function of eccentricity for $q_p = 10^{-6}$ and $q_d=10^{-4}$.  Also plotted is $T_\mathrm{migr}$ of \eq{Tmigr} for a constant value of $\gamma_\mathrm{tot}=10$, the one-sided migration timescale ($T_\mathrm{1s})$, and the type-I migration timescale $T_\mathrm{type-I}$ (\eqp{Ttype-I}). As can be seen from \fg{Tmigr}, self-regulated migration results in a migration timescale that is significantly shorter than \eq{Tmigr}, especially at low and modest eccentricities.  In fact, it approaches the migration rate due to the one-sided torque; for $e_h=3-4$ the self-regulated migration rate rivals that of as type-I.

\subsection{Preconditions for the self-regulated migration regime}
In the above analysis we have assumed that the planet exerts a modest feedback on the disk.  For the self-regulated migration mechanism to operate the planet mass can neither be too small (since then no feedback is present) nor too massive (too much feedback). If it is too small $\beta_\mathrm{sr}$ (\eqp{beta-2}) will be much lower than the global value of the eccentricity index $\beta$.  Thus, $|\beta_\mathrm{sr}| \gtrsim |\beta|$ is a rough estimate of the precondition for the self-regulated mechanism to become feasible.

Another assumption in the above analysis is that the jump in eccentricity gradient is smooth.  More specifically, for the derivation of \eq{Tsr} to be valid the timescale inequality relation
\begin{equation}
  \label{eq:ineq-0}
  T_\mathrm{syn} \ll T_\mathrm{migr}^\ast \ll T_\mathrm{vs}
\end{equation}
has to be obeyed. The first inequality ensures that the number of scattering during the passage of the planet ($T_\mathrm{migr}^\ast$) is $\gg$1: the eccentricity of a planetesimal when it proceeds downstream (in the frame of the planet) then increases gradually.  The second inequality indicates that the total jump in eccentricity, $\Delta e/e_0 \ll1$, is small: the planet is of a small enough size to only mildly excite the disk.

The synodical timescale can be approximated as $T_\mathrm{syn} = 4\pi/3e_0 \Omega_0$. \Eq{ineq-0} then reads
\begin{equation}
  \label{eq:ineq-1}
  \frac{4\pi}{3e_0} \ll \frac{e_0^3}{12 f_\Lambda \beta q_d q_p} \ll \frac{2e_0^5}{f_\Lambda q_p^2}.
\end{equation}
Converting to Hill eccentricities, $e_0 = e_h (q_p/3)^{1/3}$, and rearranging, \eq{ineq-1} transforms into
\begin{equation}
  \label{eq:ineq-2}
  6\pi f_\Lambda \ll Q_\mathrm{pd} e_h^{11/2} \ll e_h^6,
\end{equation}
where
\begin{equation}
  Q_\mathrm{pd} 
  = \frac{2^{1/2} q_p^{1/3}}{4 \cdot 3^{1/3} q_d^{1/2}}
  \simeq 0.25 q_p^{1/3} q_d^{-1/2}
  \label{eq:Qpd}
\end{equation}
is a combination of the planet and the disk masses (for the values of $q_d$ and $q_p$ used in \fg{Tmigr} $Q_\mathrm{pd}=0.25$). With this notation, the second inequality of \eq{ineq-0}  corresponds to 
\begin{equation}
  \label{eq:eh-ast1}
  e_h \gg Q_\mathrm{pd}^2,
\end{equation}
whereas the first inequality becomes
\begin{equation}
  e_h \gg e_h^\ast 
  = \left( \frac{6\pi f_\Lambda}{Q_\mathrm{pd}} \right)^{2/11} 
  \approx 2.7 \left( \frac{f_\Lambda}{3} \right)^{2/11} \left( \frac{q_p}{10^{-6}} \right)^{-2/33} \left( \frac{q_d}{10^{-4}} \right)^{1/11}.
  \label{eq:ehast}
\end{equation}
Both conditions must be satisfied.  Note that when $Q_\mathrm{pd}\lesssim1$ criterion \eq{eh-ast1} vanishes since $e_h$ is always larger than unity in the dispersion dominated regime.  Physically, $Q_\mathrm{pd}$ expresses the mass of the planet (the $q_p$ term) with respect to the planetesimal disk (the $q_d$ term).  When $Q_\mathrm{pd}$ is large, the planet has too much inertia to be affected by planetesimal scattering.  However, even when $Q_\mathrm{pd}\ll1$ it is required that $e_h>e_h^\ast$ to ensure a smooth power-law gradient in eccentricity -- a precondition in the derivation of \eq{Tsr}.  

\begin{figure}
  \centering
  \includegraphics[width=85mm]{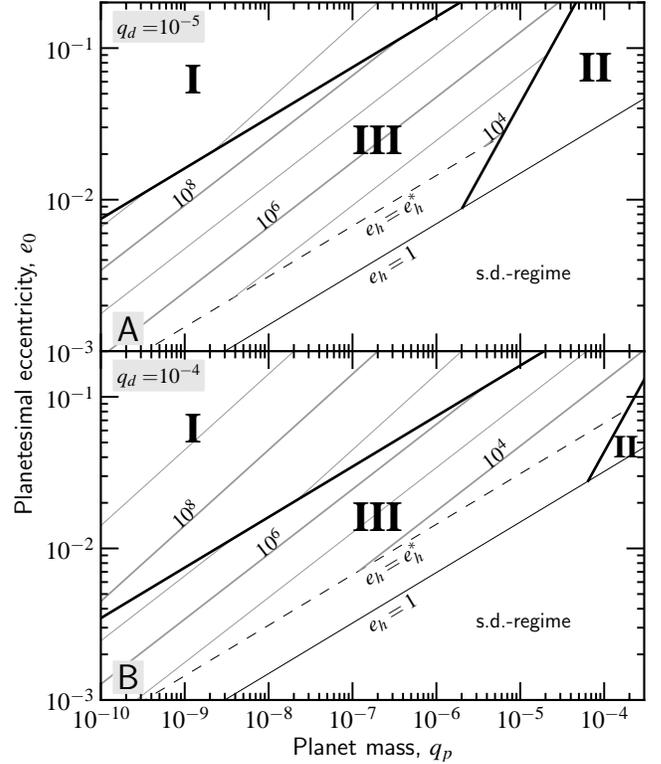}
  \caption{Classification of migration regimes for a disk mass parameter $q_d$ of $10^{-5}$ \textit{(top)} and $10^{-4}$ \textit{(bottom)} and contours \textit{(solid gray lines)} of the migration timescales in terms of $\Omega_0^{-1}$. For small planets (area I) the feedback of the planet on the disk is negligible and the planet migration timescales are given by \eq{Tmigr}. Migration in regime II is suppressed due to the large mass of the planet. Intermediate-mass planets self-regulate their migration (area III). The migration rate is then given by \eq{Tsr} as long as $e_h>e_h^\ast$ (\eq{ehast}).}
  \label{fig:Qpd}
\end{figure}

\section{Discussion}
\label{sec:diss}
\subsection{Migration regimes}
Using the results of \se{sus-migr} we can identify three regimes, where the migration behavior is qualitatively different:
\begin{enumerate}
  \item The low planet mass regime, where the planet can be regarded as a test body that does not affect the structure (surface density, eccentricity) of the disk;
  \item A large planet mass regime, where the planet has too much inertia to experience sustained migration;
  \item An intermediate regime, where the planet self-regulates its migration.
\end{enumerate}
These regimes are indicated in \fg{Qpd} by the roman literals, I, II, and III. The thick black lines denote the regime boundaries. The boundary between the first and third regime correspond to the criterion $\beta_\mathrm{sr} = 1$. The boundary between regime II and III follows from \eq{eh-ast1}. Contour lines give migration timescales $T_\mathrm{migr}$ in terms of the local orbital period.  The migration timescales of planets in regime I are those given by \eq{Tmigr}, whereas those in regime III obey \eq{Tsr} as long as $e_h>e_h^\ast$ (dashed line). In regime II, planetesimals are excited before migration becomes important. Note that the choices for the regime boundaries are somewhat flexible; in reality, the transition between the regimes will be broader than suggested by the sharp boundaries of \fg{Qpd}.

Both the migration rate as well as the position of the regime boundaries depend on the disk mass parameter $q_d$ (\eq{qdisk}). A large disk mass enlarges the fraction of the parameter space occupied by regime I; whereas a low disk mass enlarges that of regime II. The direction of the migration in regime I is defined by the exponents $\alpha$ and $\beta$, characterizing the disk-wide power-law indices of the surface density and eccentricity profiles. In regime III, the migration direction is unspecified, as the migrating planet will self-adjust $\beta$ locally to $\pm$$\beta_\mathrm{sr}$. That is, the direction depends on the history of the planet.

For example, when a planet migrates outwards in regime I due to a negative gradient in the planetesimals' eccentricity, it may cross the boundary towards regime III, where it will continue to migrate outwards at a faster pace. Similarly, a gradient in $q_d$ (a local quantity) can also trigger a regime change. Generally, we find that migration timescales in regime I are long (unless $q_d$ is large) and that fast migration occurs in the self-regulated mode. However, regime I is important in setting the initial direction of the migration.

The classification I, II, and III resembles the corresponding migration types for gas-driven migration. For type I, the feedback of the planet on the disk is negligible. As shown in \se{type-I}, the expressions for the migration timescales show very similar scalings. Type II migration is driven by diffusive motions of the gas \citep[\eg][]{GresselEtal2011} or solids. We showed in \se{diff}, however, that for PDM these timescales are generally long. On the other hand, PDM in the type-III mode covers a large region of the parameter space (\fg{Qpd}).

\subsection{Implications for $N$-body simulations}
Recently, state-of-the-art $N$-body simulations investigating the migration behavior of a single planet embedded in a massive planetesimal disk \citep{KirshEtal2009,CapobiancoEtal2011} exhibit an integration instability, \ie\ a planet migrates on timescales of the one-sided torque (\eqp{T1s}). Since in \citet{KirshEtal2009} and \citet{CapobiancoEtal2011} the interactions operate mostly in the shear-dominated regime, scatterings are strong and migration timescales short\footnote{To see this one may substitute $e_0(e_h\sim1) \simeq (q_p)^{1/3}$ in \eqp{T1s}.}.  Our work does not consider the shear-dominated regime; but similar to the above works, we discover a migration mechanism that is self-regulated.  That is, the timescale to move away from the local stirring zone of the planet ($T_\mathrm{migr}^\ast$, \eqp{Tmigr-ast}) is shorter than the viscous stirring timescale $T_\mathrm{vs}$.  As long as the conditions remain in place (most notably, the disk in planetesimals must be massive; see \fg{Qpd}) the migration does not stop.  As in \citet{KirshEtal2009}, the migration mechanism does not specify a direction (inwards or outwards); this depends on some initial perturbation.  Migration timescales down to $10^5$ times the local orbital period then seem quite viable.  

$N$-body simulations are necessarily limited in terms of their dynamic range; due to their large numbers, it is often impossible to resolve each planetesimal individually.  Therefore, the superparticle concept is often employed, in which groups of planetesimals are represented by a single $N$-body particle, which effectively amounts to increasing their gravitational mass $m$ but keeping the surface density constant \citep{KirshEtal2009,BromleyKenyon2011}. To save computational resources, one prefers a large amount of grouping (that is, a large $m$). But it is clear that a simulation involving too massive superparticles will no longer accurately reflect the physically system.  This is immediately clear in the extreme limit when $m$ approaches $M_p$, in which case a planet is scattered by the debris!  As we quantified in \se{diff} a large $m/M_p$ ratio increases the importance of diffusive motions (`noise'), and we calculated the length scale over which diffusive noise can be expected to be dominant.

Alternatively, $N$-body simulations often account for dynamical friction from a debris (small particle) component analytically.  In a recent study, \citet{LeinhardtEtal2009} studied the behavior of an $N$-body system interacting in coagulation and migration including a prescription for $\mathbf{F}_\mathrm{df}$ resulting from interaction with debris. However, in their work the debris is assumed to move on non-eccentric Keplerian orbits and the Keplerian shear is not accounted for -- both effects render the dynamical friction force artificially large. (Indeed planets are seen to migrate very rapidly inwards!). The correct procedure should follow the lines of this work; that is, one must solve for the local velocity field.  Clearly, we must generalize the calculations to include eccentric planets at arbitrary inclination and include encounters in the shear-dominated regime \citep{BromleyKenyon2011i} -- issues that will be addressed in the future.  The calculations presented in this work therefore have the potential to provide a significant boost in the accuracy of $N$-body simulations containing a debris component. 

\section{Conclusions}
In this paper, we have employed detailed analytical and numerical calculations to obtain the net torque acting on a planet due to the recoil from gravitational interactions with planetesimals in the dispersion-dominated regime.  We have included both distant and close encounters and obtained the net migration rate by summing the torques from the interior and the exterior disks. We list our conclusions:
\begin{enumerate}
  \item While the magnitude of the migration rate is primarily determined by the local values of the surface density ($\Sigma_0$) and eccentricity ($e_0$), the direction is given by the local gradient in these quantities ($\alpha$ and $\beta$) and by the Coulomb factor $f_\Lambda$ (\fg{Itot}).  Usually, the contribution from close encounters will determine the migration direction, unless $f_\Lambda$ (and by implication $e$) are low.
  \item The expressions for the migration timescale \eq{Tmigr} due to planetesimal scattering display similarities to type-I migration (\eq{Ttype-I}), if one replaces the disk mass in planetesimals by that of the gas and the eccentricity by $c_s/V_k$.  Since the disk mass in solids is lower, the planetesimal-driven migration timescale is generally longer.
  \item Under certain conditions a much faster migration mode (rivaling that of type-I) is obtained when the feedback of the planet on the disk is accounted for. The planet then self-regulates the value of the eccentricity gradient to $\beta_\mathrm{sr}$, which is a function of the local physical parameters (\eq{beta-2}).  Generally, $|\beta_\mathrm{sr}| \gg 1$ and migration is quite rapid (\eq{Tmigr}).
  \item As function of the dimensionless disk mass $q_d$ (\eq{qdisk}), planet mass $q_p=M_p/M_\star$, and planetesimal eccentricity $e$, we have identified three migration regimes (\fg{Qpd}) representing: (I) low mass planets, for which disk excitation is negligible; (II) high-mass planets, too massive to migrate significantly; and (III) intermediate-mass planets, which exert a mild feedback on the disk and migrate in the self-regulated mode.

\end{enumerate}

\acknowledgments
C.W.O.\ acknowledges valuable discussions with Marco Spaans and Eugene Chiang. Support for this work was provided by NASA through Hubble Fellowship grant \#HST-HF-51294.01-A awarded by the Space Telescope Science Institute, which is operated by the Association of Universities for Research in Astronomy, Inc., for NASA, under contract NAS 5-26555.

\bibliographystyle{apj}
\bibliography{ads}

\appendix

\section{Torque density and migration rate for distant encounters}
\label{app:Idist}
A planet in a gas or particular disks exerts a torque on the disk at discrete positions (resonances).  These resonances are identified by integer numbers $l,m$, which refer to the corresponding Fourier modes. For given $m$, the strongest resonance occurs where $l=m$, for which the resonance condition (a relation between $m$ and the distance to the planet $b$) reads:
\begin{equation}
  \label{eq:res-con}
  \Omega(r) \left( 1 +\frac{\varepsilon}{m} \right) = \Omega_0,
\end{equation}
where $\varepsilon = \pm1 = \sgn{b}$ refers to an inner or outer Lindblad resonance and $\varepsilon=0$ refers to a co-rotation resonance. 

Here we consider Lindblad resonances. For $m\gg1$ the resonances start to overlap and a continuum treatment is possible. Therefore, a torque density, $d\Gamma/dr$, can be defined \citep{GoldreichTremaine1980}. We follow the notation of \citep{Ward1997}, who defines:
\begin{equation}
  \label{eq:T-lm}
  \frac{d\Gamma}{dr}
  = 
  \varepsilon \frac{2q_p^2 \Sigma a_0^4 \Omega_0^2}{r} \left[ m^4 \psi^2_m \left( \frac{\Omega_0}{\kappa[r]} \right)^2 \right]_\mathrm{res}.
\end{equation}
where $\psi_m$ is an expression that involves the computation of Laplace coefficients, and $\kappa$ is the epicyclic frequency at $r$. It is important that the RHS of \eq{T-lm} is evaluated at resonance, whose condition is given by \eq{res-con}.  

We redefine $b$ such that it becomes dimensionless, $b \rightarrow b/a_0 \ll 1$, and expand \eq{res-con} with respect to $b$:
\begin{equation}
  m = 
  \frac{\epsilon}{(1+b)^{3/2} -1}
  \approx \frac{2}{3|b|} \left( 1 -\frac{b}{4} \right),
\end{equation}
where $\epsilon = \sgn{b}$. \citet{Ward1997} has obtained the function $\psi_m^2$ to first order in $m$ (or $b$). He finds
\begin{equation}
  \label{eq:psi2}
  \psi^2(b)
  = \left[2K_0(2/3) +K_1(2/3)\right]^2 \left[ 1+ \frac{b}{2} \left\{ \frac{ K_0(2/3) +5K_1(2/3)}{2K_0(2/3) +K_1(2/3)}  \right\} \right]
  \approx 6.35(1+1.255b)
\end{equation}
Thus, the four $b$-dependent terms in \eq{T-lm} expand as follows: 
$\psi^2(b)$ as in \eq{psi2}; 
$1/r \approx (1+b)/a_0$; $m^4 \approx (2/3b)^4(1-b)$; 
and the epicycle frequency as $(\Omega_0/\kappa)^2 \approx 1 +3b$ since $\kappa(r)=\Omega(r)$ in a Keplerian disk. Collecting these terms gives the torque density to first order in $b$:
\begin{equation}
  \label{eq:dTdb}
  \frac{d\Gamma}{db} 
  \approx \sgn{b} \frac{2^5 q_p^2 \Sigma a_0^4 \Omega_0^2}{3^4 b^4} 
    \left[2K_0(2/3) +K_1(2/3)\right]^2
    \left( 1 +2.255b \right),
\end{equation}
where $K_\nu(x)$ is the modified Bessel function of second kind of order $\nu$ (Note again that in this equation $b$ is nondimensional). To zeroth order ($\Sigma=\Sigma_0$) therefore:
\begin{equation}
  \label{eq:dTdb-dist0}
  \left( \frac{d\Gamma}{db} \right)_0 
  = \sgn{b} (M_p a_0^2 \Omega_0^2) \frac{2^5 q_p q_d }{3^4 b^4} \left[2K_0(2/3) +K_1(2/3)\right]^2
  \approx 2.51 (M_p a_0^2 \Omega_0^2) \sgn{b} \frac{q_p q_d}{b^4},
\end{equation}
where we substituted $q_p = M_p/M_\star$ and $q_d = a_0^2\Sigma_0/M_\star$.

The torque from the disk on the planet has the opposite sign as \eq{dTdb}. When we consider one only side of the disk, the leading term will not vanish. For example, the torque from the outer disk on the planet is:
\begin{equation}
  \frac{\Gamma_\mathrm{1s-di}}{M_p (a_0\Omega_0)^2}
  \equiv -\int_e^\infty db\ \left( \frac{dT_0}{db} \right)_0
  = \mp \frac{2^5 q_p q_d}{3^5 e_0^3} \left[2K_0\besselarg +K_1\besselarg \right]^2
  \approx \mp 0.836 M_p \frac{q_d q_p}{e_0^3},
\end{equation}
(where the sign is positive for an interior disk). This expression is consistent with the migration rate derived in \eq{dadt-dist}.

To zeroth order, the integration over both sides of the disk vanishes due to symmetry of \eq{dTdb-dist0} with respect to $b$.  In the next order expansion of \eq{dTdb}, however, nonzero contributions originate due to:
\begin{enumerate}
  \item The curvature term in \eq{dTdb}; and the gradient in surface density $\alpha$. These terms render the integrand odd ($\propto$$1/b^3$) and together contribute a factor
    \begin{equation}
      - 2\int_e^{\infty} db \left( \frac{d\Gamma}{db} \right)_0 (2.26+\alpha)b 
      = -3e_0 (2.26 +\alpha) \left| \Gamma_\mathrm{1s-di} \right|
    \end{equation}
  \item The gradient in eccentricity. This causes the transition between close and distant encounters to shift by a value $b = \beta e_0$ (see \fg{disk}): from $-e_0 \le b \le e_0$ (when $\beta=0$) to $-e_0+\beta e_0^2 \le e_0 +\beta e_0^2$. The corresponding net torque due to this shift is the zeroth order torque density at $b=e_0$ times twice the width of this shift
    \begin{equation}
      \int_{e_0(1-\beta)}^{e_0(1+\beta)} db \left( \frac{dT}{db} \right)_0
      \approx 2\beta e_0^2 \times \left( \frac{d\Gamma}{db} \right)_0[b=e_0]
      = 6\beta e_0 |\Gamma_\mathrm{1s-di}|
    \end{equation}
\end{enumerate}
Summing the two terms gives the leading term of the torque on the planet when accounting for both sides of the disk:
\begin{equation}
  \Gamma_\mathrm{2s-di}
  = -3e_0(2.26 +\alpha -2\beta) |\Gamma_{1s-di}|
  = M_p (a_0 \Omega_0)^2 q_d q_p \frac{-5.66 -2.51(\alpha -2\beta)}{e_0^2}. 
  \label{eq:Gam-2s}
\end{equation}

\section{Calculation of integrals for close encounters}
\label{app:Iclose}
Let us write the \eq{V-close-1} in nondimensional form:
\begin{eqnarray}
  \label{eq:Iclose-app}
  \frac{\Gamma_{\mathrm{cl},\phi}}{a_0^2 M_p \Omega_0^2}
  &=& \frac{G_N^2 M_p \Sigma_0}{(a_0\Omega_0)^4} \times \left[ 2\pi f_\Lambda  \int_{a_\mathrm{in}}^{a_\mathrm{ou}} \frac{\dd a}{a_0}\ a_0^2\tilde{P}_{Rz} \left( \frac{a}{a_0} \right)^{1+\alpha} \frac{v_\phi V_{K0}^2}{v^3} \right] \\
  &\equiv& q_d q_p \times I_\mathrm{cl}(\alpha,\beta),
  \nonumber
\end{eqnarray}
where we used $(G_N M_\star)^2 = (a_0^3\Omega_0^2)^2$. The integral in the square brackets is defined $I_\mathrm{cl}(\alpha,\beta)$ and must be computed. 

The integration is over the range in semi-axes $a$ where planetesimals are able to cross the planet's orbit $a_0$.  In the above $v=|\mathbf{v}|$ is the velocity of an unperturbed body at $a=a_0$ relative to the circularly-moving planet and $v_\phi$ the azimuthal component of $\mathbf{v}$.  Both $\mathbf{v}$ and $\tilde{P}_{Rz}$, the probability density of planetesimals at $a_0$ (see below), are functions of the semi-major axis $a$ of the planetesimals.  In the following sections we will obtain expressions for $\mathbf{v}$ and $P_{Rz}$, respectively.  For aesthetic purposes we will express these quantities, as well as the other terms in \eq{Iclose-app}, as function of $\theta$ -- the true anomaly of the Kepler orbit at the point where it intersects the planet -- rather than $a$.  A perturbation analysis in $e_0$ then allows us to compute \eq{Iclose-app} analytically.

\subsection{Expressions resulting from the Kepler orbit}
\begin{figure}
  \centering
  \includegraphics[width=14cm]{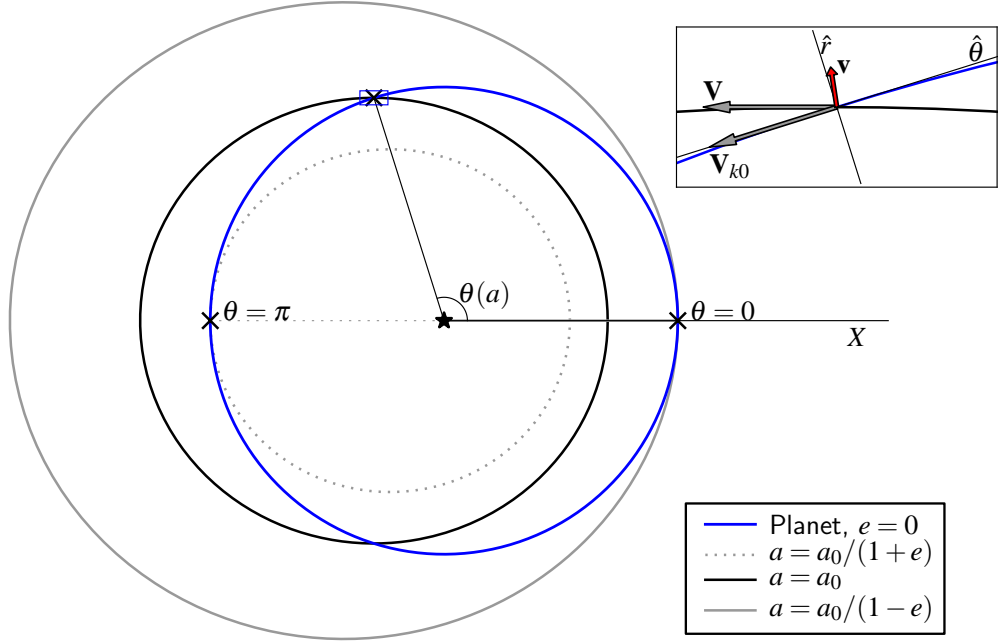}
  \caption{Sketch of sample trajectories in the orbital plane.  A planet moves on a circular orbit of radius $a_0$ (\textit{blue circle}) and interacts with planetesimals that are characterized by semi-major axis $a$ and eccentricity $e=0.3$.  For fixed $e$ the range in $a$ that crosses the planet's orbit is $a_0/(1+e) \le a \le a_0/(1-e)$.  The periapsis   ($\theta=0$) of the orbits are, for clarity, situated on the positive $X$ axis. The planet interacts with the outermost planetesimal swarm (\textit{gray, solid circle}) at \textit{their} periapsis ($\theta=0$) and with the innermost swarm (\textit{gray, dotted circle}) at \textit{their} apoapsis ($\theta=\pi$). For planetesimals of intermediate $a$, \eg\ those of $a=a_0$ (\textit{black circle}), $\theta$ is given by \eq{d-a-th}.  At the interaction point (see \textit{inset}) the planet moves along the $\hat{\theta}$ direction at the Keplerian velocity $V_{k0}$. The components of the planetesimal velocity $\mathbf{V}$ are given by \eqs{Vth}{Vr}. For the calculation of the dynamical friction force it is the relative velocity $\mathbf{v} = \mathbf{V} -\mathbf{V}_{k0}$ that matters (\textit{red arrow}).\\ \\ \\ }
  \label{fig:sketch}
\end{figure}

A body traveling on a Kepler orbit obeys the relation
\begin{equation}
  r = \frac{a(1-e^2)}{1+e\cos\theta},
\end{equation}
where $r$ is the radial coordinate in the orbital plane of the particle, $a$ the semi-major axis, $e$ the eccentricity, and $\theta$ the true anomaly specifying the instantaneous position of the particle.  This orbital plane is inclined by an angle $i$ with respect to the equatorial plane -- the plane in which the planet moves.  The particle intersects the planet's circular orbit at $r=a_0$, or at a true anomaly $\theta$ that obeys the relation:
\begin{equation}
  a = a_\theta = a_0 \frac{1+e\cos\theta}{(1-e^2)}.
  \label{eq:d-a-th}
\end{equation}
Thus, there is a one-to-one relation between the semi-major axis $a$ from which the planetesimal originates and the true anomaly $\theta$ at which it crosses the planet at $a_0$.  We will use \eq{d-a-th} to switch the integration variable to $\theta$:
\begin{equation}
  \left| \frac{da}{d\theta} \right| = \frac{a_0e \sin\theta}{1-e^2}.
  \label{eq:dadt}
\end{equation}
Next, we express the Keplerian velocities also as function of $\theta$. Kepler's second law gives the azimuthal velocity, $V_\theta=l/r=\sqrt{a(1-e^2)\mu}/r$ with $l$ the angular momentum and $\mu=G_N(M_\star+M_p)\approx G_N M_\star$. The radial velocity can be obtained from energy conservation.  We evaluate these expressions at the interaction point, \ie\ at $r=a_0$ and $a/a_0$ given by \eq{d-a-th}:
\footnote{The expression for the radial velocity expression is perhaps difficult to see at first glance.  Energy conservation gives
\begin{equation}
  \frac{1}{2}(V_r^2+V_\theta^2) 
  = -\frac{\mu}{2a} +\frac{\mu}{a_0} 
  = \frac{\mu}{a_0} \left( 1-\frac{a_0}{2a} \right) 
  = \frac{V_{k0}^2}{2} \left( \frac{1+2e\cos\theta +e^2}{1+e\cos\theta} \right)
\end{equation}
where we used \eq{d-a-th}.  Then, inserting \eq{Vth} for $V_\theta$:
\begin{equation}
  V_r^2 = V_{k0}^2 \left(  \frac{1+2e\cos\theta +e^2}{1+e\cos\theta} - \frac{(1+e\cos\theta)^2}{1+e\cos\theta} \right)
\end{equation}
and \eq{Vr} is retrieved.
}
\begin{eqnarray}
  \label{eq:V-plts}
    \label{eq:Vth}
    V_\theta 
    &=& \sqrt{\frac{G_N M_\star (1-e^2)}{a_0} \frac{a}{a_0}} 
    = V_{k0} \sqrt{1+e\cos \theta}; \\
    V_r 
    &=& V_{k0} \frac{e \sin \theta}{\sqrt{1+e\cos \theta}},
    \label{eq:Vr}
\end{eqnarray}
where $V_{k0}=\sqrt{\mu/a_0}$ is the orbital velocity at the interaction point (the Kepler velocity at which the planet moves). In disk (\ie\ cylindrical) coordinates ($R,\phi,z$) the velocities at the interaction point $(R,z)=(a_0,0)$ at arbitrary $\phi$ become:
\begin{eqnarray}
  \label{eq:V3D}
    V_R &=& V_r; \\
    V_\phi &=& V_\theta \cos i; \\
    V_z &=& V_\theta \sin i;
\end{eqnarray}
and the relative velocity vector, written in terms of $\theta$, reads:
\begin{equation}
  \label{eq:vrel}
  \mathbf{v} = \mathbf{V} -\mathbf{V}_p =
  \left( \begin{array}{l}
    V_R \\ V_\phi - V_{k0} \\ V_z
  \end{array} \right)
   = V_{k0}
  \left( \begin{array}{l}
    \displaystyle
    \frac{e \sin \theta}{\sqrt{1+e\cos \theta}} \\[4mm]
    \displaystyle
    \cos i \sqrt{1+e\cos \theta} -1 \\[1mm]
    \displaystyle
    \sin i \sqrt{1+e\cos \theta} \\
  \end{array} \right).
\end{equation}

\subsection{The projection operator $\tilde{P}_\mathrm{Rz}$}
The orbit of a planetesimal is additionally determined by $\omega$, the angle of periapsis.  Transforming from $(r,\theta)$ to Cartesian coordinates in the equatorial plane gives \citep{Beutler2005}:
\begin{eqnarray}
    \label{eq:x-cyl}
    x &=& r_\theta \cos(\omega+\theta); \\
    y &=& r_\theta \sin(\omega+\theta) \cos i; \\
    \label{eq:z-cyl}
    z &=& r_\theta \sin (\omega + \theta) \sin i;
\end{eqnarray}
where we have assumed, without loss of generality, that the line of nodes is directed along the $x$-axis.  From these equations we obtain the projected radius on the equatorial plane $R$:
\begin{equation}
  \label{eq:R-cyl}
  R = \sqrt{r^2 -z^2} = r\sqrt{1-\sin^2(\omega+\theta) \sin^2 i}.
\end{equation}
For the problem considered here, $R$ and $z$ are the principal variables. The density function $P_{Rz}$ is defined such that $P_{Rz} \dd R \dd z$ gives the probability of finding the particle within $[R,R+\dd R; z,z+\dd z]$.  To obtain $P_{Rz}$, we assume that the phase angles $t$ (mean anomaly) and $\omega$ are randomly distributed, $P_{t,\omega}=\Omega_a/(2\pi)^2$, where $\Omega_a$ denotes the orbital frequency corresponding to semi-major axis $a$. Converting  variables then gives:
\begin{equation}
  P_{Rz} = \left|\frac{\partial(t,\omega)}{\partial(R,z)}\right| P_{t,\omega},
  \label{eq:PRz}
\end{equation}
where $\partial(t,\omega)/\partial(R,z)$ is the Jacobian of the transformation.  

We may proceed to use Kepler's equation to relate the mean anomaly $t$ to $\theta$.  Here, however, we consider a specific case where $P_{Rz}$ is only evaluated at $(R,z)=(a_0,0)$.  Let this density be denoted $\tilde{P}_{Rz}$. From \eq{z-cyl} it can be seen that $z=0$ corresponds to either $\omega=-\theta$ or $\omega=\pi-\theta$.  Therefore, $\tilde{P}_{Rz}$ is a function of one variable only, say $\theta$.  Formally, we define
\begin{equation}
  \label{eq:Ptrz}
  \tilde{P}_{Rz} = P_{Rz}(z=0) = \int \dd\omega\ \left| \frac{\partial(t,\omega)}{\partial(R,z)} \right| P_{t,\omega} \left[ \delta(\omega=-\theta) + \delta(\omega=\pi-\theta) \right],
\end{equation}
where $\delta(x)$ is the Dirac delta function.  Using \eq{R-cyl} and \eqsto{x-cyl}{z-cyl} and anticipating that the matrix elements of the Jacobian will be evaluated at $r_\theta = a_0$ and $\omega = -\theta$ or $\omega=\pi-\theta$:
\begin{eqnarray}
    \frac{\partial z}{\partial \omega} 
    &=& r_\theta \cos(\omega+\theta) \sin i  \rightarrow a_0 \sin i \\
    \frac{\partial z}{\partial \theta} 
    &=& r_\theta \cos(\omega+\theta) \sin i + \frac{\partial r}{\partial\theta}\sin(\omega+\theta) \sin i \rightarrow a_0 \sin i \\
    \frac{\partial R}{\partial \omega} 
    &=& \frac{z (\partial z/\partial \omega)}{\sqrt{r^2+z^2}} \rightarrow 0 \\
    \frac{\partial R}{\partial \theta} 
    &=& \frac{r(\partial r/\partial \theta) -z(\partial z/\partial \theta)}{\sqrt{r^2-z^2}} \rightarrow \frac{\partial r}{\partial \theta}\quad \Rightarrow\quad \frac{\partial R}{\partial t} = V_r
\end{eqnarray}
where `$\rightarrow$' indicates we have evaluated the matrix elements at $z=0$ (or $\omega=-\theta)$ and $R=a_0$.
The Jacobian, evaluated at $(R,z)=(a_0,0)$ then reduces to
\begin{equation}
  \left| \frac{\partial(t,\omega)}{\partial(R,z)} \right|_{R=a_0; z=0} 
  = \left| \frac{\partial R}{\partial t} \frac{\partial z}{\partial \omega} \right|_{R=a_0; z=0}
  = \frac{1}{V_r a_0 \sin i}
\end{equation}
an expression that only involves $\theta$ as a variable.  Substituting, we obtain for the density function \eq{Ptrz}
\begin{equation}
  \label{eq:Prz}
  \tilde{P}_{Rz}(\theta)= \frac{\Omega_a(\theta)}{2\pi^2 V_r a_0 \sin i}.
\end{equation}
Substituting \eq{Vr} for $V_r$ and $\Omega_a = \Omega_0 (a_\theta/a_0)^{-3/2}$ we have
\begin{equation}
  \label{eq:a02-Prz}
  a_0^2 \tilde{P}_{Rz}(\theta) = \frac{\sqrt{1+e\cos\theta}}{2\pi^2 e \sin \theta \sin i} \left( \frac{a_\theta}{a_0} \right)^{-3/2}.
\end{equation}
\Eq{Prz} can be validated numerically, for example by distributing particles on a Kepler orbit characterized by fixed orbital elements $a$, $e$, and $i$ but random $\omega$ and $t$.  The midplane number density corresponding to the reference radius $a_0$, $n_\mathrm{mid}(a_0[\theta])$, is then $n_\mathrm{mid}(a_0) = 2\tilde{P}_{Rz}/2\pi a_0$. The additional factor of 2 takes care of the fact that there are two $\theta$-solutions for a given $a_0$.

\subsection{Integral evaluations}
Using \eq{dadt} to transform coordinates to $\theta$ and \eq{a02-Prz} for $a_0^2 \tilde{P}_{Rz}$, $I_\mathrm{cl}(\alpha,\beta)$ (see \eqp{Iclose-app}) can be expressed solely as a function of $\theta$:
\begin{equation}
  \label{eq:Iclose-app2}
  I_\mathrm{cl} = 2\pi f_\Lambda \int d\theta\ \frac{\sqrt{1+e\cos\theta}}{2\pi^2 \sin i (1-e^2)} \left( \frac{a_\theta}{a_0} \right)^{\alpha-1/2} \frac{v_\phi V_{k0}^2}{v^3},
\end{equation}
where $a_\theta$, $v_\phi$ and $v$ are given by \eqs{d-a-th}{vrel}.  Furthermore, $e$ and $i$ are also functions of $\theta$ by virtue of the gradient (\eqp{sige0}). Since \eq{Iclose-app2} cannot be solved algebraically, we expand it in $e$ to obtain a closed-form solution. Concerning the inclination, we will consider two cases: \sumi\ $i=e/2\ll1$; and \sumii\ $i \ll e \ll 1$.

\subsubsection{The equilibrium solution, $i=e/2$}
Substituting $i=e/2$ in \eq{Iclose-app2} and approximate $e$, assuming $e\ll1$:
\begin{equation}
  \label{eq:e-th}
  e(\theta) \approx e_0 + \beta e_0^2 \cos\theta.
\end{equation}
Next, we expand \eq{Iclose-app2} around $e_0$, accounting only for terms of ${\cal O}(e_0^{-3})$ and ${\cal O}(e_0^{-2})$. The integrand of \eq{Iclose-app2} then becomes:
\begin{equation}
 \label{eq:integ-Icl}
  f_\Lambda \left\{ 
  \frac{44 \cos (\theta )-12 \cos (3 \theta )}{2\pi \left(\cos ^2(\theta )+4 \sin ^2(\theta)+1\right)^{5/2}}e_0^{-3}
  +\frac{22 \alpha-66\beta+2 (8 \alpha-24\beta+1) \cos (2 \theta )-3 (2 \alpha-6\beta+3) \cos (4 \theta )-9}{2\pi  \left(\cos ^2(\theta )+4 \sin ^2(\theta )+1\right)^{5/2}}e_0^{-2}
 \right\}.
\end{equation}
The $e_0^{-3}$ term is symmetric and evaluates to 0 when the full range of $\theta$ is considered.  When we integrate only over one side of the disk (\eg\ $-\pi/2 \le \theta \le \pi/2$) this term determines the one-sided torque:
\begin{equation}
  I_\mathrm{1s-cl}^{(i=e/2)}
  = \frac{8f_\Lambda}{\sqrt{5}\pi e_0^3} 
  \approx \frac{1.14f_\Lambda}{e_0^3}.
  \label{eq:I1s}
\end{equation}
The $e_0^{-2}$ term in \eq{integ-Icl} does not vanish after integration over $0\le \theta \le 2\pi$ and evaluates to
\begin{equation}
  \label{eq:Iclose-app3}
  \label{eq:Iclose-app4}
  I_\mathrm{2s-cl}^{(i=e/2)} =
  \frac{4 \sqrt{2} \left(E\left(-\frac{3}{2}\right) (12 \alpha -36 \beta
   +5)+K\left(-\frac{3}{2}\right) (-12 \alpha +36 \beta -11)\right)}{9 \pi e_0^2}
   f_\Lambda
\approx \frac{-0.66 +1.97(\alpha-3\beta)}{e_0^2}
   f_\Lambda.
\end{equation}
where $E(x)$ and $K(x)$ are complete elliptic integrals of the first and second kind:
\begin{equation}
  K(x) = \int_0^{\pi/2} \frac{d\theta}{\sqrt{1-x\sin^2\theta}};\qquad 
  E(x) = \int_0^{\pi/2} d\theta \sqrt{1-x\sin^2\theta}.
\end{equation}

\subsubsection{The thin disk case, $i\ll e$}
Next we consider a case of a thin disk $(i\ll e$).  A thin disk is applicable when the planet interacts with the planetesimals in the shear-dominated regime.  Note that we have assumed in the main text that the dispersion-dominated regime applies, for which $i/e\approx0.5$ is expected, but a situation where $i\ll e$ can still emerge as a transient state \citep{RafikovSlepian2010}.

We consider the situation where the inclination also obeys a power-law:
\begin{equation}
  i(a) = i_0 \left( \frac{a}{a_0} \right)^\delta \approx i_0 + \delta i_0 e_0 \cos\theta.
  \label{eq:ia}
\end{equation}
We follow the same procedure as above, expanding $I_\mathrm{cl}$ first in terms of $i_0$ and then in terms of $e_0$. Subsequently, integration over $\theta$ gives:
\begin{eqnarray}
  \label{eq:Iclose-thin} 
  I_\mathrm{1s-cl}^{(i\ll e)} 
  &=& \frac{4f_\Lambda}{\pi i_0 e_0^2}
  \approx \frac{1.27f_\Lambda}{i_0 e_0^2} \\
  \nonumber
    I_\mathrm{2s-cl}^{(i\ll e)} &\approx&
    \frac{12 \left[-2 K(-3)-K\left(\frac{3}{4}\right)+2 E(-3)+4 E\left(\frac{3}{4}\right)\right] (\alpha
   -2 \beta -\delta ) 
   -14\left[2 K(-3)+K\left(\frac{3}{4}\right)\right]
   +44 E\left(\frac{3}{4}\right)+22 E(-3)}{9 \pi e_0 i_0} f_\Lambda
  \\ &\approx&
  \frac{1.63 + 2.28 (\alpha-2\beta-\delta)}{i_0e_0} f_\Lambda .
\end{eqnarray}
for the one- and two-sided integrals, respectively.

\subsubsection{General case}
\label{app:general}
Within the limits of our assumptions (inclinations and eccentricities larger than the Hill eccentricity, \etc) we consider an even more general case. After inserting \eq{e-th} and \eq{ia} for $e(\theta)$ and $i(\theta)$, respectively, we define $i_0 = \zeta e_0$ and expand $I_\mathrm{cl}$ in terms of $e_0$. The resulting expression (a function of $\zeta$ and $\theta$) can be integrated. This procedure gives:
\begin{eqnarray}
  I_\mathrm{1s} &=& \frac{4f_\Lambda}{\pi e_0^3 \sqrt{\zeta ^2+1} \left(4 \zeta ^3+\zeta \right)} \\
  I_\mathrm{2s} &=&
  \frac{4 E\left(-\frac{3}{4 \zeta ^2+1}\right) A_1
  +8A_2 \left(\zeta ^2+1\right) K\left(-\frac{3}{4 \zeta ^2+1}\right) }{9 \pi e_0^2 \sqrt{4 \zeta ^2+1}
   \left(4 \zeta ^5+5 \zeta ^3+\zeta \right)}f_\Lambda
\end{eqnarray}
where
\begin{eqnarray}
  A_1 &=& 4 \zeta ^4 (12 \alpha -12 \beta -24 \delta -1)+\zeta ^2 (60 \alpha -36 \beta -144 \delta +7)+12 \alpha -24 \beta -12 \delta +11 \\
  A_2 &=& 8 \zeta ^2 (-3 \alpha +3 \beta +6 \delta -2)-6 \alpha +12 \beta +6 \delta -7
\end{eqnarray}
It can be verified that these formula recover the expressions for the equilibrium regime (where $\zeta=1/2$ and $\delta=\beta$) and the $i\ll e$ regime (where $\zeta \ll 1$), respectively.

\section{Calculation for the diffusion integrals (close encounters)}
\label{app:diff}
\Eq{Ddelvy2} gives the rate of change in $\Delta v_\phi^2$:
\begin{equation}
  \label{eq:Ddelvy2-app}
  \Diffs{\phi} = 2\pi G^2_N m^2 \left\{ \frac{nv_\phi^2}{v^3} [3g_\Lambda-f_\Lambda] +\frac{n}{v}[f_\Lambda-g_\Lambda] \right\}.
\end{equation}
The number density $n$, and $v$ are functions of disk radius $a$ and \eq{Ddelvy2-app} must accordingly be converted in an integration over $a$.  Following a similar procedure as described in \se{close-main}, we write:
\begin{equation}
  D[(\Delta v_\phi)^2] = 2\pi G^2_N m  \int \dd{a}\ \tilde{P}_{Rz} \Sigma(a) \left( \frac{a}{a_0} \right) \left\{  (3g_\Lambda -f_\Lambda)\frac{v_\phi^2}{v^3} + \frac{f_\Lambda -g_\Lambda}{v} \right\}.
  \label{eq:Ddelvy2-app2}
\end{equation}
After inserting \eq{sige0} for $\Sigma(a)$ we split the integral, $\Diffs{\phi} \equiv {\cal D}_0 (I_1 + I_2)$, with ${\cal D}_0$ the dimensional part given by
\begin{equation}
  {\cal D}_0 = 
  \frac{G^2_N m \Sigma_0}{a_0^2 \Omega_0} = 
  \frac{a_0^2 \Sigma_0}{M_\star} \frac{m}{M_\star} a_0^2 \Omega_0^3
\end{equation}
and $I_1$ and $I_2$ dimensionless integrals given by
\begin{eqnarray}
  \label{eq:I1I2}
  I_1 &=& 2\pi(3g_\Lambda -f_\Lambda) \int da'\ \tilde{P}_{Rz}'  \left( \frac{a}{a_0} \right)^{1+\alpha} \frac{v_\phi'^2}{v'^3}\\
  I_2 &=& 2\pi(f_\Lambda -g_\Lambda)  \int da'\ \tilde{P}_{Rz}'  \left( \frac{a}{a_0} \right)^{1+\alpha} \frac{1}{v'}
\end{eqnarray}
where primes denote normalized quantities.

The procedure to evaluate integrals \eq{I1I2} is the same as in \app{Iclose}.  First, we change variables to $\theta$ via \eqs{d-a-th}{dadt}. Then we insert $\tilde{P}_{Rz}(\theta)$ and the velocity field, $\mathbf{v}(\theta)$, as given by \eqs{a02-Prz}{vrel}, respectively. We further assume equilibrium, $i=e/2$, insert $e(\theta)$ (\eqp{e-th}) and expand $I_1, I_2$ in $e_0$, keeping only the highest order term.  Subsequently, we derive:
\begin{eqnarray}
  I_1 &= &
  -\frac{8 \sqrt{2} (f_\Lambda-3 g_\Lambda)
   \left(E\left(-\frac{3}{2}\right)-K\left(-\frac{3}{2}\right)\right)}{3
   e_0^2 \pi}
   \approx \frac{0.98 (3g_\Lambda-f_\Lambda )}{e_0^2} \\
  I_2 &=&
  \frac{8 \sqrt{2} (f_\Lambda-g_\Lambda)
    K\left(-\frac{3}{2}\right)}{e_0^2 \pi}
    \approx \frac{4.4 (f_\Lambda -g_\Lambda)}{e_0^2} \\
  I_1 + I_2 &\approx&
  \frac{3.5 f_\Lambda -1.5 g_\Lambda}{e_0^2}.
\end{eqnarray}
Note that there is no dependence on $\alpha$ or $\beta$ as the highest-order terms do not cancel.
\end{document}